\documentclass[a4paper,fleqn]{cas-sc}
\usepackage{bm}

\makeatletter
\def\Hy@Warning#1{}
\makeatother

\usepackage{longtable}

\def\tsc#1{\csdef{#1}{\textsc{\lowercase{#1}}\xspace}}
\tsc{WGM}
\tsc{QE}
\tsc{EP}
\tsc{PMS}
\tsc{BEC}
\tsc{DE}
\begin{document}
\let\WriteBookmarks\relax
\def\floatpagepagefraction{1}
\def\textpagefraction{.001}
\shorttitle{Finite element solver for a thermodynamically consistent electrolyte model}
\shortauthors{Habscheid et~al.}
%\begin{frontmatter}

\title[mode = title]{A finite element solver for a thermodynamically consistent electrolyte model}

%%%%%%%%%%%%%%%%%%%%%
%%%%%%%%%%%%%%%%%%%%%%
\author{Jan Habscheid}[%
	orcid=0009-0005-0040-1562%
]
\ead{jan.habscheid@rwth-aachen.de}
\author{Satyvir Singh}[%
	orcid=0000-0001-6669-5296%
]
\cormark[1]
\ead{singh@acom.rwth-aachen.de}
\cortext[cor1]{Corresponding author}
\author{Lambert Theisen}[%
	orcid=0000-0001-5460-5425%
]
\ead{lambert.theisen@rwth-aachen.de}
\author{Stefanie Braun}
\ead{braun@acom.rwth-aachen.de}
\author{Manuel Torrilhon}[%
	orcid=0000-0003-0008-2061%
]
\ead{mt@acom.rwth-aachen.de}
\address{Institute for Applied and Computational Mathematics, RWTH Aachen University, Germany}

%%%%%%%%%%%%%%%%%%%%%%%%%%%%%%%%%%%%%%%%%%%%%%%%%%%%%%%%%%%%%%%%%%%%%

\begin{abstract}
In this study, we present a finite element solver for a thermodynamically consistent electrolyte model that accurately captures multicomponent ionic transport by incorporating key physical phenomena such as steric effects, solvation, and pressure coupling. The model is rooted in the principles of non-equilibrium thermodynamics and strictly enforces mass conservation, charge neutrality, and entropy production. It extends beyond classical frameworks like the Nernst--Planck system by employing modified partial mass balances, the electrostatic Poisson equation, and a momentum balance expressed in terms of electrostatic potential, atomic fractions, and pressure, thereby enhancing numerical stability and physical consistency. Implemented using the FEniCSx platform, the solver efficiently handles one- and two-dimensional problems with varied boundary conditions and demonstrates excellent convergence behavior and robustness. Validation against benchmark problems confirms its improved physical fidelity, particularly in regimes characterized by high ionic concentrations and strong electrochemical gradients. Simulation results reveal critical electrolyte phenomena, including electric double layer formation, rectification behavior, and the effects of solvation number, Debye length, and compressibility. The solver's modular variational formulation facilitates its extension to complex electrochemical systems involving multiple ionic species with asymmetric valences. We publicly provide the documented and validated solver framework.
\end{abstract}


\begingroup
	\hfuzz=200pt  % avoid overfull hbox warning
	\maketitle
\endgroup

%%%%%%%%%%%%%%%%%%%%%%%%%%%%%%%%%%%%%%%%%%%%%%%%%%%%%%%%%%%%%%%%%%%%%%%%%%%%%%
%%%%%%%%%%%%%%%%%%%%%%%%%%%%%%%%%%%%%%%%%%%%%%%%%%%%%%%%%%%%%%%%%%%%%%%%%%%%%%

%%%%%%%%%%%%%%%%%%%%%%%%%%%%%%%%%%%%%%%%%%%%%%%%%%%%%%%%%%%
%%%%%%%%%%%%%%%%%%%%%%%%%%%%%%%%%%%%%%%%%%%%%%%%%%%%%%%%%%
\section{Introduction}\label{Sec:1}

The worldwide shift from fossil fuels to renewable energy has significantly amplified the need for efficient, scalable energy storage systems that are essential for supporting both stationary grid infrastructure and mobile applications~\cite{yang2011electrochemical}. As lithium-ion batteries approach their theoretical performance limits, significant research efforts have shifted toward next-generation alternatives such as metal-sulfur and metal-air batteries, polymer electrolyte systems, and solid-state batteries---each offering the potential for higher energy densities and enhanced safety~\cite{janek2016solid,lotsch2017relevance,placke2017lithium,yue2016all}. Simultaneously, the growing global need for clean and accessible water, driven by rising population, industrialization, and agricultural demands, has catalyzed the development of advanced water treatment technologies. Electrically driven processes like electrodialysis and capacitive deionization are gaining attention for their promise of energy-efficient, scalable desalination, in contrast to traditional thermal or pressure-driven methods~\cite{anderson2010capacitive,suss2015water}.

At the heart of both energy storage and water purification technologies lies a common enabler: electrolyte solutions. These ionic media are fundamental to a wide range of scientific and engineering domains, including electrochemical systems~\cite{xia2017electrolytes}, biological environments~\cite{wright2007introduction}, chemical manufacturing~\cite{aseyev2014electrolytes}, and environmental remediation~\cite{anderko2002electrolyte}. The performance, stability, and efficiency of these systems depend critically on the behavior of electrolytes, particularly at charged interfaces where processes such as ion exchange, electrode reactions, membrane selectivity, and double-layer formation occur. Capturing this behavior accurately requires mathematical and computational models that seamlessly integrate principles of thermodynamics, electrostatics, and transport phenomena.

%$$$$$$$$$$$$$$$$$$$$$$$$$$$$$$$$$$$$$$$$$$$
\begin{figure}
	\centering
	\includegraphics[width=0.8\textwidth]{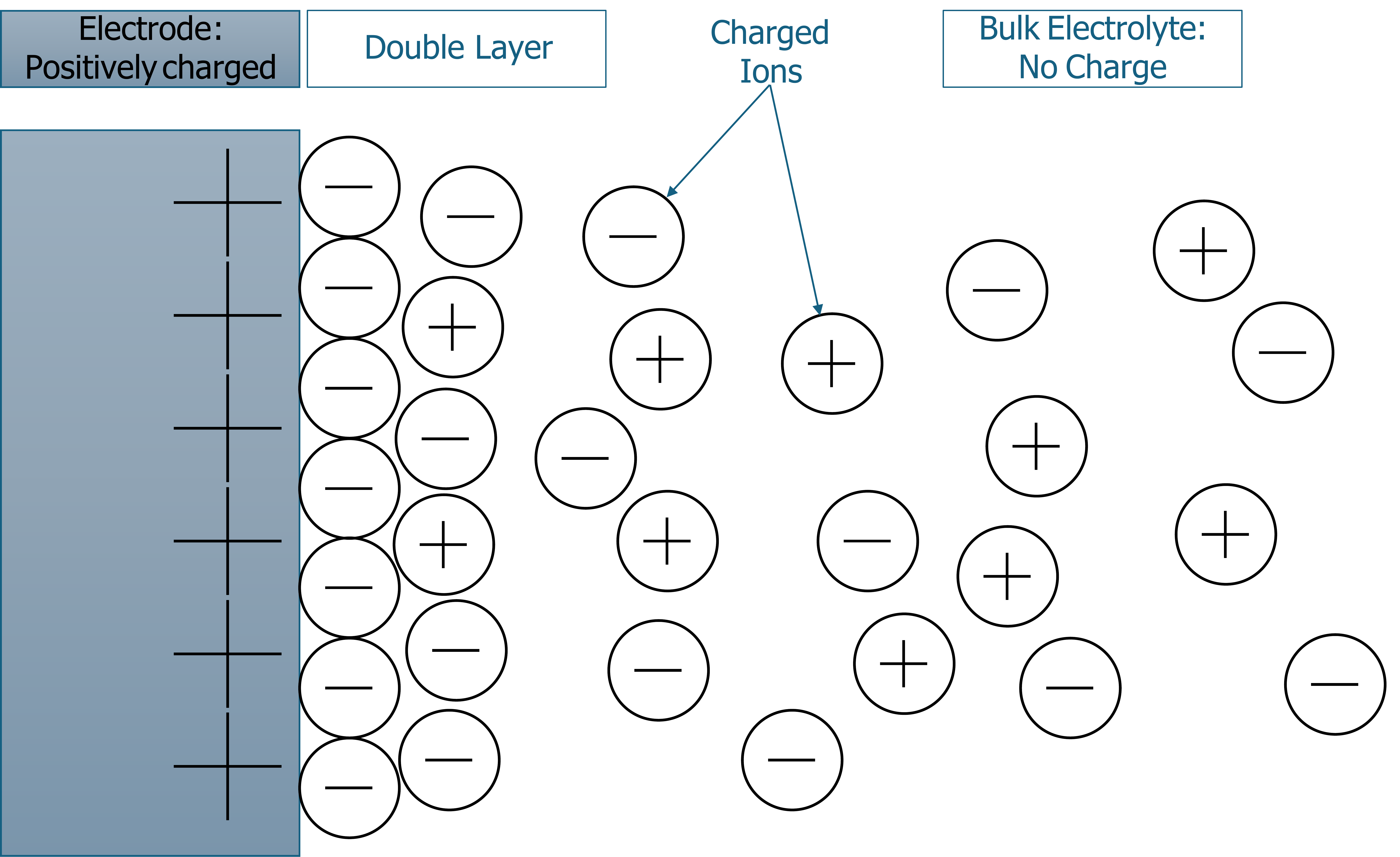}
	\caption{Schematic of the electrical double layer (EDL) near a positively charged electrode. Negatively charged ions accumulate near the electrode surface, forming the compact layer, while the bulk electrolyte remains electrically neutral with a uniform distribution of positive and negative ions.}\label{Fig:1}
\end{figure}
%$$$$$$$$$$$$$$$$$$$$$$$$$$$$$$$$$$$$$$$$$$$

To emphasize the significance of electrolyte behavior in energy storage devices, Figure~\ref{Fig:1} shows a schematic of the electrical double layer (EDL) near a positively charged electrode. When a voltage is applied in a battery, ions in the electrolyte migrate toward the oppositely charged electrode, forming a structured layer. Negatively charged ions accumulate near the electrode surface, creating a Stern layer, while the surrounding ions form a diffuse layer. This process influences charge storage, ion transport, and electrochemical reactions, directly impacting battery performance, efficiency, and lifespan. Understanding the formation and behavior of the EDL is essential for optimizing electrode materials and improving energy storage technologies. The concept of EDL is particularly important in lithium-ion batteries, supercapacitors, and fuel cells, where ion distribution and electrostatic interactions play a crucial role in determining energy storage efficiency and power density. Enhancing the understanding of EDL dynamics can lead to the development of more efficient energy storage devices with improved charge retention and faster charge/discharge cycles.

%%%%%%%%%%

%%%%%%%%%%%%%%%%%%%%%%%%%%%%%%%%%%% electrolyte models
%%%%%%%%
Over the years, several mesoscopic frameworks have been proposed to model the complex interaction between ion transport and electrochemical phenomena in electrolyte systems. One of the most widely used is the classical Nernst--Planck (NP) model, which describes ion fluxes as driven by concentration gradients and electric fields through diffusion and electromigration~\cite{Nernst1888,Nernst1889,Planck1890one,Planck1890two}. Although effective for dilute electrolytes under near-equilibrium conditions, the NP model overlooks important physical effects such as finite size effects, ion--ion correlations, and thermodynamic coupling. These limitations can lead to nonphysical results, including negative concentrations and energy inconsistency, especially in systems with high ionic strength or strong fields~\cite{kilic2007steric,dreyer2013overcoming,horng2012pnp}. The Poisson--Nernst--Planck (PNP) model extends the NP formulation by coupling it with the Poisson equation, enabling the electrostatic potential to evolve self-consistently with local charge distributions~\cite{jerome2012analysis}. This framework is widely used to simulate ion transport in nanopores, membranes, and biological ion channels. However, the standard PNP model shares many of the NP model's shortcomings and does not adequately capture steric and finite size effects, or non-ideal mixing---key factors in concentrated or strongly coupled systems~\cite{ottinger2005beyond,grmela1997dynamics}. The size-modified Poisson--Nernst--Planck (SMPNP) formulation addresses such limitations through additional entropic terms~\cite{luPoissonNernstPlanckEquationsSimulating2011}.

%%%%%%%%%%%%%%%%%%%%%%%%%%%%%%%%%%%%%%%%%%%%%%%555555
%%%%%%%%%%%%%%%%%%%%%%%%%%%%%%%%%%%%%%%%%%%%%%%%%%%%
To overcome the limitations inherent in classical ion transport models---such as the neglect of ion--ion interactions, finite size effects, and lack of thermodynamic consistency---several advanced formulations have been developed. One notable class of models replaces the traditional flux--force relations with those derived from chemical potentials, resulting in volume-exclusion models that incorporate non-idealities through concentration-dependent chemical potentials~\cite{sparnaay1958corrections,Kornyshev1981Conductivity,Mansoori1971equilibrium,Tresset2008generalized}. These approaches provide improved physical fidelity by accounting for steric and finite size effects. However, they also introduce analytical and computational complexities, particularly in systems near thermodynamic equilibrium or under strong coupling conditions.

Building on these advancements, thermodynamically consistent models have emerged as a robust and physically grounded framework for modeling ionic transport. Rooted in non-equilibrium thermodynamics and often formulated via variational principles, these models incorporate essential physical constraints such as electrochemical potentials, energy dissipation, and entropy production, offering a predictive and stable description of ion dynamics in systems where chemical, electrical, and mechanical interactions are strongly coupled. Such formulations not only address the shortcomings of classical models but also ensure compliance with the second law of thermodynamics, making them suitable for a broad range of applications.

Several important contributions have advanced this framework. De Groot et al.~\cite{de2013non} emphasized the role of finite size effects in establishing thermodynamically consistent relationships between chemical potential and concentration. Dreyer et al.~\cite{dreyer2013overcoming} introduced models that rigorously couple mechanical forces with ionic diffusion, successfully addressing multiple deficiencies of the classical NP model. Roubíček~\cite{roubivcek2006incompressible} proposed a comprehensive coupled model integrating the Poisson equation, heat equation, NP equations, and the Navier--Stokes system. However, this formulation did not fully conform to the second law of thermodynamics, particularly when coupling the $N-1$ diffusion fluxes.

To improve numerical treatment, Prohl and Schmuck~\cite{prohl2010convergent} developed and analyzed two fully discrete, convergent schemes for the incompressible Navier--Stokes/Poisson--Nernst--Planck (NS/PNP) system. In the context of solid electrolytes, Braun et al.~\cite{braun2015thermodynamically} presented a thermodynamically consistent model for ionic transport and derived a semi-analytical solution for a one-dimensional stationary case, highlighting how system parameters influence the formation of space-charge layers under applied voltage. Fuhrmann~\cite{Fuhrmann2015} extended the classical PNP framework to include modified formulations with enhanced physical realism and improved numerical properties. In a follow-up work, Fuhrmann~\cite{fuhrmann2016numerical} introduced a fully thermodynamically consistent model for an isothermal, incompressible ionic mixture in mechanical equilibrium, incorporating finite size effects and varying solvation numbers to provide a more comprehensive description of transport phenomena in electrolyte systems. Recently, Ankur et al.~\cite{Ankur2025electrolyte} proposed a finite element approach for a modified NS/PNP model, which incorporates a bulk modulus parameter to enable seamless transition between compressible and incompressible regimes. Their study demonstrates the applicability of the model across a range of conditions, including variations in temperature and fluid compressibility.

%%%%%%%%%%%%%%%%%%%%%%%%%%%%%%
%%%%%%%%
From a computational standpoint, the literature addressing the numerical approximation of modified Poisson--Nernst--Planck (PNP) and Navier--Stokes/PNP (NS/PNP) systems with thermodynamic consistency remains relatively limited. In contrast, there is a substantial body of work dedicated to the classical PNP equations and their coupling with fluid flow through the NS equations~\cite{Prohl2009, Jerome1991, Yang2013, Ray2012, Prohl2010, Schmuck2009, Schmuck2011}. Among various numerical methods, the finite element method (FEM) has proven to be a versatile and effective approach for solving these coupled systems.
Several advancements have extended classical models to improve physical realism. For example, He and Sun~\cite{Mingyan} introduced a modified NS/PNP formulation in which the standard Poisson equation is replaced by a fourth-order elliptic equation to better capture electrostatic interactions. Sun et al.~\cite{Sun2016} developed a fully nonlinear Crank--Nicolson FEM scheme for the PNP equations, while Gao and He~\cite{Gao2017} proposed a linearized FEM with optimal error estimates. Despite these contributions, the strong coupling between electrochemical transport and hydrodynamics still presents analytical and numerical challenges, particularly in establishing well-posedness and ensuring stability for generalized or thermodynamically consistent formulations.
The existence and uniqueness of solutions for the steady-state PNP system have been explored in early theoretical works~\cite{Jerome1985, Park1997}, with further extensions to coupled NS/PNP systems~\cite{Jerome2002}. To enhance numerical robustness and flexibility, various alternative discretization strategies have been proposed, mostly relying on finite element discretizations using both conforming and discontinuous Galerkin methods~\cite{Gao2018, He2017, He2018, Kim2022, Linga2020, liu2017free, liu2022positivity, Xie2020}. Recently, Ding and Zhou introduced a positivity-preserving, energy-stable second-order scheme for modified Poisson--Nernst--Planck equations~\cite{dingSecondorderPositiveUnconditional2024}.
Nevertheless, despite notable progress in electrolyte modeling, the development of efficient, stable, and thermodynamically consistent numerical solvers for modified electrolyte systems remains a significant challenge. Classical approaches often fail to preserve key physical properties such as energy dissipation, entropy production, and charge neutrality, especially under nonlinear or strongly coupled conditions. These limitations motivate the present study, which aims to address these gaps through a robust computational framework.

%%%%%%%%%%%%%%%%%%%%%% Motivation and scope
% \paragraph{Outline and contribution}
In this manuscript, we present a FEniCS-based finite element solver for a thermodynamically consistent electrolyte model~\cite{dreyer2013overcoming} that is derived via a variational formulation grounded in non-equilibrium thermodynamics. It enforces mass conservation, charge neutrality, and entropy consistency, and is validated against experimental measurements~\cite{dreyer2018bulk,landstorferTheoryStructureMetalelectrolyte2016}.
By leveraging the FEM flexibility, the proposed solver is capable of handling complex geometries, heterogeneous boundary conditions, and multi-physics couplings. We validate the numerical convergence, stability, and accuracy of the solver against other computational benchmarks.
Furthermore, we explore the influence of key model parameters, including ionic diffusivities, spatial variations in dielectric permittivity, and externally applied fields, on electrolyte dynamics.
Together, these contributions yield a robust and versatile computational tool, which we release publicly to advance the understanding and design of electrochemical devices.
\par
The remainder of this paper is organized as follows. In Section~\ref{Sec:2}, we describe the governing equations and their derivation based on non-equilibrium thermodynamics. Section~\ref{Sec:3} presents the details of the finite element implementation, including the discretization strategies used. Section~\ref{Sec:4} provides validation and benchmark results for the developed solver and model performance, while Section~\ref{Sec:5} provides a detailed presentation of further results and discussion. Finally, Section~\ref{Sec:6} summarizes our findings and suggests potential future directions for extending this work.

%%%%%%%%%%%%%%%%%%%%%%%%%%%%%%%%%%%%%%%%%%%%%%%%%%%%%%%%%%%
%%%%%%%%%%%%%%%%%%%%%%%%%%%%%%%%%%%%%%%%%%%%%%%%%%%%%%%%%%
\section{Mathematical formulation for electrolyte model}\label{Sec:2}

%%%%%%%%%%%%%%%%%%%%%%%%%%%%%%%%%%%%%%%%%%%%%%%%%%%%%%%%%%
\subsection{Notation and preliminaries}\label{Sec:2.1}

In this section, we introduce the notation and symbols used throughout
this work. Let $\Omega\subset\mathbb{R}^{3}$ be a three-dimensional
spatial domain occupied by a liquid mixture composed of $N$ different
ionic species. Each species is indexed by $\alpha\in\{1,2,\ldots,N\}$
and is characterized by its partial mass density $\rho_{\alpha}$,
velocity $\boldsymbol{v}_{\alpha}$, atomic mass $m_{\alpha}$, number
density $n_{\alpha}$, and charge number $z_{\alpha}$.

The mixture is influenced by a spatially and temporally varying electric
field $\boldsymbol{E}$ and has a uniform background temperature $T$.
Consequently, the scalar quantities $\rho_{\alpha}$ and $n_{\alpha}$,
and vector fields such as $\boldsymbol{v}_{\alpha}$ and $\boldsymbol{E}$
depend on both space and time:
\begin{equation}
[\rho_{\alpha},n_{\alpha}]:\Omega\times\mathbb{R}^{+}\rightarrow\mathbb{R}^{+},\quad[\boldsymbol{v}_{\alpha},\boldsymbol{E}]:\Omega\times\mathbb{R}^{+}\rightarrow\mathbb{R}^{3}.\label{Eq:1}
\end{equation}
The partial mass density is also given as
\begin{equation}
\rho_{\alpha}=m_{\alpha}n_{\alpha},\label{Eq:2}
\end{equation}
and the total mass density $\rho$ and total momentum $\rho\boldsymbol{v}$
of the mixture are given by
\begin{equation}
\rho=\sum_{\alpha=1}^{N}\rho_{\alpha},\quad\rho\boldsymbol{v}=\sum_{\alpha=1}^{N}\rho_{\alpha}\boldsymbol{v}_{\alpha}.\label{Eq:3}
\end{equation}
The total number density is $n=\sum_{\alpha=1}^{N}n_{\alpha}$, and
the atomic (mole) fraction of each species is
\begin{equation}
y_{\alpha}=\frac{n_{\alpha}}{n},\quad\sum_{\alpha=1}^{N}y_{\alpha}=1,\quad0\leq y_{\alpha}\leq1,\label{Eq:4}
\end{equation}
while the barycentric (mass-averaged) velocity $\boldsymbol{v}$ and the
diffusion velocity $\boldsymbol{u}_{\alpha}$ are defined as:
\begin{equation}
\boldsymbol{v}=\frac{1}{\rho}\sum_{\alpha=1}^{N}\rho_{\alpha}\boldsymbol{v}_{\alpha},\quad\boldsymbol{u}_{\alpha}=\boldsymbol{v}_{\alpha}-\boldsymbol{v}.\label{Eq:5}
\end{equation}
The diffusion flux of species $\alpha$ is given by
\begin{equation}
\boldsymbol{J}_{\alpha}=\rho_{\alpha}\boldsymbol{u}_{\alpha},\quad\sum_{\alpha=1}^{N}\boldsymbol{J}_{\alpha}=0.\label{Eq:6}
\end{equation}
The total charge density $n^{e}$ consists of free and polarization
contributions
\begin{equation}
n^{e}=n^{F}+n^{P},\label{Eq:7}
\end{equation}
where the free contributions are
\begin{equation}
n^{F}=\sum_{\alpha=1}^{N}z_{\alpha}e_{0}n_{\alpha},\label{Eq:8}
\end{equation}
with $e_{0}$ being the elementary charge. The polarization contributions
are defined as
\begin{equation}
n^{P}=-\text{div}(\boldsymbol{P}),\label{Eq:9}
\end{equation}
through the material polarization vector $\boldsymbol{P}$.

\subsection{Thermodynamically consistent electrolyte model}\label{Sec:2.2}

To address the limitations of classical ion transport models, a thermodynamically
consistent electrolyte model is formulated based on non-equilibrium
thermodynamics~\cite{dreyer2013overcoming}. The governing equations
respect conservation of mass, momentum, and electrostatic constraints
while ensuring non-negative entropy production. The model represents finite-size (steric) saturation through ideal mixing and, optionally, solvation-induced specific volumes; explicit hard-core ionic diameters and nonlocal excluded-volume correlations are not included.

The unclosed equations read
\begin{equation}
\begin{aligned}\partial_{t}\rho_{\alpha}+\text{\text{div}}(\rho_{\alpha}\boldsymbol{v}+\boldsymbol{J}_{\alpha}) & =0,\quad\alpha\in\{1,\dots,N-1\}\\
\partial_{t}\rho+\text{div}(\rho\boldsymbol{v}) & =0,\\
\partial_{t}(\rho\boldsymbol{v})+\text{div}(\rho\boldsymbol{v}\otimes\boldsymbol{v}+\boldsymbol{\sigma}) & =n^{e}\boldsymbol{E},\\
\epsilon_{0}\text{div}(\boldsymbol{E}) & =n^{e},
\end{aligned}
\label{eq:unclosed}
\end{equation}
which require constitutive relations for the stress tensor $\boldsymbol{\sigma}$
and mass fluxes $\boldsymbol{J}_{\alpha}$, as well as, for the polarization
field $\boldsymbol{P}$ as it occurs in the polarized part of the total charge density. In~\eqref{eq:unclosed}, there are $N-1$ partial mass balances because their sum yields the total mass balance.
The constitutive theory defines the free energy of the system as
\begin{align}\label{eq:free_energy_function}
\rho\psi{\left(\rho_{\alpha},\boldsymbol{E}\right)} & =\sum_{\alpha=1}^{N}\rho_{\alpha}\psi_{\alpha}^{\mathrm{\text{ref}}}+\rho\psi_{\text{M}}+\rho\psi_{\text{E}}+\rho\psi_{\text{P}},
\end{align}
with an isotropic, elastic deformation part $\psi_{\text{M}}$, an entropic
mixing part $\psi_{\text{E}}$ and a part due to dielectric polarization
$\psi_{\text{P}}$ which are given by
\begin{align}
\begin{aligned} & \rho\psi_{\text{M}}=\left(K-p^{\text{ref}}\right)\left(1-{\textstyle \frac{n}{n^{\text{ref}}}}\right)+K\frac{n}{n^{\text{ref}}}\ln{\left({\textstyle \frac{n}{n^{\text{ref}}}}\right)},\\
 & \rho\psi_{\text{E}}=nkT\sum_{\alpha=1}^{N}y_{\alpha}\ln(y_{\alpha}),\\
 & \rho\psi_{\text{P}}=-\frac{1}{2}\varepsilon_{0}\chi|\boldsymbol{E}|^{2},
\end{aligned}
\end{align}
where the superscript $()^{\text{ref}}$ indicates a constant reference
value and $k$ is the Boltzmann constant. The material parameters
are the bulk modulus $K$ and the dielectric susceptibility $\chi$
which are both considered constant. This formulation incorporates
compressibility effects through the bulk modulus, and captures non-ideal
mixing via the logarithmic term. Then thermodynamic theory (with~\eqref{eq:free_energy_function} and~\eqref{Eq:17}) yields
\begin{align}
\mu_{\alpha} & =\frac{\partial(\rho\psi)}{\partial\rho_{\alpha}}=g_\alpha+\frac{kT}{m_{\alpha}}\ln(y_{\alpha}),\label{eq:chem_pot}
\\
\boldsymbol{P} & =-\frac{\partial(\rho\psi)}{\partial\boldsymbol{E}}=\varepsilon_{0}\chi\boldsymbol{E},\label{eq:p_propto_e}\\
\boldsymbol{\sigma} & =\left(\rho\psi-\sum_{\alpha=1}^{N}\rho_{\alpha}\mu_{\alpha}\right)\boldsymbol{I}+\boldsymbol{E}\otimes\boldsymbol{P}=p\boldsymbol{I}+\varepsilon_{0}\chi{\left(\boldsymbol{E}\otimes\boldsymbol{E}-\frac{1}{2}|\boldsymbol{E}|^{2}\boldsymbol{I}\right)},
\end{align}
which includes the chemical potential $\mu_{\alpha}$. In~\eqref{eq:chem_pot}, the specific Gibbs energy $g_{\alpha}$ is given by
\begin{align}\label{eq:SpecificGibbs}
	g_\alpha & =g_{\alpha}^{\text{ref}}+\frac{K}{n^{\text{ref}}m_{\alpha}}\ln{\left(\frac{n}{n^{\text{ref}}}\right)},
\end{align}
while
\begin{align}\label{eq:pressure}
p & =p^{\text{ref}}+K{\left(\frac{n}{n^{\text{ref}}}-1\right)},
\end{align}
denotes the total pressure.

The mass fluxes are obtained from the entropy production $\xi$, ensuring
the second law of thermodynamics:
\begin{equation}
\xi=-\sum_{\alpha=1}^{N}\boldsymbol{J}_{\alpha}\cdot\Bigg(\nabla\bigg(\frac{\mu_{\alpha}}{T}\bigg)-\frac{1}{T}\frac{z_{\alpha}}{m_{\alpha}}\boldsymbol{E}\Bigg)\geq0.
\end{equation}
Following Dreyer et al.~\cite{dreyer2013overcoming}, we consider
$N-1$ independent diffusion fluxes given by
\begin{equation}\label{eq:diff_fluxes}
\boldsymbol{J}_{\alpha}=-\sum_{\beta=1}^{N-1}\boldsymbol{M}_{\alpha\beta}\Bigg(\nabla\bigg(\frac{\mu_{\beta}-\mu_{N}}{T}\bigg)-\frac{1}{T}\Big(\frac{z_{\beta}}{m_{\beta}}-\frac{z_{N}}{m_{N}}\Big)\boldsymbol{E}\Bigg),\quad\alpha\in\{1,\dots,N-1\},
\end{equation}
where the material coefficient $\boldsymbol{M}_{\alpha\beta}$ is a symmetric, positive-definite mobility
matrix while $\boldsymbol{J}_{N}$ follows from~\eqref{Eq:6}.

Typically, the electric field is derived from a potential $\boldsymbol{E}=-\nabla\varphi$.
After replacing $n^{e}$ with $n^{F} + n^{P}$, using~\eqref{Eq:9} and~\eqref{eq:p_propto_e} as well as inserting the expression
for the stress tensor and polarization the complete system consists
of $N-1$ independent partial mass balances, the total mass and momentum
balances, and Poisson's equation for the electrostatic
potentials, and can be rewritten
\begin{equation}
\begin{aligned}\partial_{t}(m_{\alpha}n_{\alpha})+\text{\text{div}}(m_{\alpha}n_{\alpha}\boldsymbol{v}+\boldsymbol{J}_{\alpha}) & =0,\quad\alpha\in\{1,\dots,N-1\},\\
\partial_{t}\rho+\text{div}(\rho\boldsymbol{v}) & =0,\\
\partial_{t}(\rho\boldsymbol{v})+\text{div}(\rho\boldsymbol{v}\otimes\boldsymbol{v})+\nabla p & =-n^{F}\nabla\varphi,\\
-\epsilon_{0}(1+\chi)\Delta\varphi & =n^{F}.
\end{aligned}
\label{Eq:10}
\end{equation}
The model ensures thermodynamic consistency across all species and
accommodates strong electrochemical interactions, making it suitable
for modeling concentrated electrolytes and highly coupled electrochemical-mechanical
systems.

%%%%%%%%%%%%%%%%%%%%%%%%%%%%%%%%%%%%%%%%%%%%%%%%%%%%%%%%%%
%%%%%%%%%%%%%%%%%%%%%%%%%%%%%%%%%%%%%%%%%%%%%%%%%%%%%%%%%%
\subsection{Model simplification and reformulation}\label{Sec:2.3}

To facilitate the analytical and numerical treatment of the thermodynamically consistent electrolyte model, several simplifying assumptions are introduced.
First, we assume that the system is in mechanical (hydrostatic) equilibrium, which allows us to set the barycentric velocity $\boldsymbol{v} = 0$ without loss of generality, consistent with prior work~\cite{fuhrmann2016numerical}. Additionally, we consider only the stationary (steady-state) regime, implying that all time derivatives vanish.
We assume that the kinetic (mobility) matrix is diagonal ($\boldsymbol{M}_{\alpha\beta}=\operatorname{diag}(\{M_\alpha\})$) and the system is isothermal, with constant temperature $T_0$. We neglect the influence of gravitational forces and allow for the presence of chemical reactions. Furthermore, the neutral solvent (species with index $N$) is assumed to be electroneutral, i.e., $z_N = 0$.

Under these assumptions, the total mass balance equation from the original system~\eqref{Eq:10} becomes redundant and is omitted. The system reduces to the following equations
%%%%%%%%%%%%%%%%%%%%%%%%%%%
%==================================
\begin{equation}
\begin{aligned}
	\label{Eq:15}
	\text{div}(\boldsymbol{J}_{\alpha}) & =0,\qquad\qquad\alpha\in\{1,\dots,N-1\}, \\
	\nabla p & =-n^{F}\nabla\varphi, \\
	-\epsilon_{0}(1+\chi)\Delta\varphi & =n^{F},
\end{aligned}
\end{equation}
%===================================
where $\nabla p =-n^{F}\nabla\varphi$ reflects the simplified momentum balance, indicating that the electric potential acts as a driving force for pressure in regions of nonzero space charge $n^F$. The term $-\epsilon_{0}(1+\chi)\Delta\varphi =n^{F}$  is the Poisson equation governing the electrostatic potential $\varphi$.

The diffusion fluxes from~\eqref{eq:diff_fluxes} simplify and are given by the scaled form with $z_N = 0$ as
%===================================
\begin{equation}
	\label{Eq:16}
	\boldsymbol{J}_{\alpha} =-M_{\alpha}\Bigg(\nabla\Big(\frac{\mu_{\alpha}-\mu_{N}}{T_{0}}\Big)+\frac{1}{T_{0}}\frac{z_{\alpha}}{m_{\alpha}}\nabla\varphi\Bigg),\hphantom{\big(-\frac{z_{N}}{m_{N}}\big)\quad}\ensuremath{\alpha}\ensuremath{\in}\{1,\dots,N-1\}.
\end{equation}
%===================================
Using~\eqref{eq:SpecificGibbs} and~\eqref{eq:pressure}, we can express the specific Gibbs energy $g_\alpha$ and the total number density $n$ in terms of $p$ as
%===================================
\begin{equation}
	\label{Eq:17}
	g_\alpha = g_\alpha^{\text{ref}} + \frac{K}{m_\alpha n^{\text{ref}}} \ln{\left(1 + \frac{p - p^{\text{ref}}}{K} \right)}, \quad
	n = n^{\text{ref}} {\left(1 + \frac{p - p^{\text{ref}}}{K} \right)},
\end{equation}
%===================================
Although the primary variables of the model are the electric potential $\varphi$, pressure $p$, and species number densities $n_\alpha$, it is often more convenient to reformulate the system in terms of atomic (mole) fractions $y_\alpha$. Accordingly, the state variables are redefined as $(\varphi, y_\alpha, p)$ for $\alpha \in \{1, \dots, N-1\}$.

%%%%%%%%%%%%%%%%%%%%%%%%%%%%%%%%%%%%%%%%%%%%%%%%%%%%%%%%%%
%%%%%%%%%%%%%%%%%%%%%%%%%%%%%%%%%%%%%%%%%%%%%%%%%%%%%%%%%%
\subsection{Comparison to classical Nernst--Planck}\label{Sec:2.4}

The diffusion fluxes derived from the thermodynamically consistent formulation include contributions from the gradients of both the species' chemical potentials and the electrostatic potential, along with the chemical potential of the neutral solvent. This comprehensive structure ensures thermodynamic consistency and proper coupling between all species.

In contrast, the classical Nernst--Planck (NP) model describes the flux of each ionic species $\alpha \in \{1, \dots, N\}$ as
%---------------------------------
\begin{equation}
		\label{Eq:18}
	\boldsymbol{J}_{\alpha}=-M_{\alpha}^{\text{NP}}\Big(k\nabla n_{\alpha}+z_{\alpha}e_{0}n_{\alpha}\nabla\varphi\Big),\quad\alpha\in\{1,\dots,N\},
\end{equation}
%---------------------------------
where $M^{\text{NP}}_\alpha > 0$ is the NP mobility, $k$ is Boltzmann's constant, $e_0$ the elementary charge, and $\varphi$ the electrostatic potential. This formulation accounts for diffusive and electromigrative driving forces.

However, the classical NP model introduces several critical simplifications:
\begin{itemize}
	\item It treats all $N$ fluxes independently, ignoring the constraint that only $N-1$ of them are linearly independent due to mass conservation.
	\item It neglects the influence of the neutral solvent's chemical potential, which is essential for maintaining a consistent coupling across species and ensuring realistic behavior near boundaries.
\end{itemize}
These oversights can result in non-physical behavior---such as artificial concentration gradients or flux imbalances---particularly in boundary layers or in systems with strong electrochemical coupling.
Thus, while the NP model is effective in dilute and near-equilibrium systems, it lacks the thermodynamic rigor necessary for accurately modeling concentrated electrolytes, non-ideal mixtures, or strong field regimes. The thermodynamically consistent model proposed in this work addresses these deficiencies by incorporating chemical potential gradients and enforcing proper constraints on the fluxes.

%%%%%%%%%%%%%%%%%%%%%%%%%%%%%%%%%%%%%%%%%%%%%%%%%%%%%
%%%%%%%%%%%%%%%%%%%%%%%%%%%%%%%%%%%%%%%%%%%%%%%%%%%%%%%%%%
%%%%%%%%%%%%%%%%%%%%%%%%%%%%%%%%%%%%%%%%%%%%%%%%%%%%%%%%%%
\subsection{Physical model specifications}\label{Sec:2.5}

We proceed with discussing model enhancements like solvation effects and properties like the incompressible limit. Additionally, dimensionless quantities are introduced to facilitate generic numerical simulations.

%%%%%%%%%%%%%%%%%%%%%%%%%%%%%%%%%%%%%%%%%%%%%%%%%%%%%%%%%%
\subsubsection{Solvation effect}\label{Sec:2.5.1}

Solvation refers to the microscopic electrostatic interactions between ionic species and surrounding solvent molecules, resulting in the formation of solvation shells. These shells consist of finite clusters of solvent molecules that arrange themselves around an ion due to local dipolar interactions~\cite{dreyer2013overcoming}. The extent of solvation is characterized by the solvation number $\kappa_\alpha$, which denotes the average number of solvent molecules associated with each ion of species $\alpha$. A higher solvation number indicates a more pronounced solvation effect.
% By definition, the neutral solvent itself does not undergo solvation, leading to $\kappa_N = 0$.
The specific volume $V_\alpha$ of species $\alpha$ can be approximated based on its solvation number using the relation:
%---------------------------------
\begin{equation}
	\label{Eq:19}
	V_\alpha = (\kappa_\alpha + 1) V_N \quad \forall \alpha \in \{1, \ldots, N-1\},
\end{equation}
%---------------------------------
where $V_N$ is the specific volume of the neutral solvent. This relation captures the volume exclusion caused by solvation and is a simple yet effective way to incorporate solvation effects into the model~\cite{dreyer2018bulk}.

Although the original thermodynamically consistent formulation in~\cite{dreyer2013overcoming} does not include solvation explicitly, it can be integrated seamlessly into the framework by modifying the expression for the specific Gibbs energy. With the solvation effect included, the specific Gibbs energy becomes:
%---------------------------------
\begin{equation}
	\begin{aligned}
	\label{Eq:20}
	g_{\alpha} & =g_{\alpha}^{\text{ref}}+(\kappa_{\alpha}+1)\frac{K}{m_{\alpha}n^{\text{ref}}}\ln\Bigg(1+\frac{p-p^{\text{ref}}}{K}\Bigg),
\end{aligned}
\end{equation}
%---------------------------------
which reduces to the original one when setting $\kappa_{\alpha}=0$. Note that the neutral solvent does not support solvation, by definition, resulting in $\kappa_N=0$.
\par
The modification above is an intentionally minimal, first-order treatment of solvation in the spirit of~\cite{dreyer2018bulk}. It assumes that solvated ions occupy a larger partial specific volume than both the neutral solvent and the unsolvated ions, which restricts the applicability to solute-solvent mixtures containing a neutral solvent. Alternatively, more comprehensive approaches like~\cite{dreyer2014mixture,fuhrmann2016numerical} derive more advanced models with an additional reformulation also for pressure and total number density. In the present work, we adopt the simpler approximation to keep the finite element formulation compact and robust, while leaving extensions for future study.

%%%%%%%%%%%%%%%%%%%%%%%%%%%%%%%%%%%%%%%%%%%%%%%%%%%%%%%%%%
\subsubsection{Incompressible limit}\label{Sec:2.5.2}

In the incompressible limit, the electrolyte mixture is assumed to have an infinitely large bulk modulus, i.e., \( K \to \infty \), indicating that the mixture resists volume changes under pressure. Applying this limit to the expression for the specific Gibbs energy in~\eqref{Eq:20}, we obtain
%--------------------------------
\begin{equation}
	\label{Eq:21}
	\lim_{K \to \infty} g_\alpha = \lim_{K \to \infty} \left( g_\alpha^{\text{ref}} + \frac{(\kappa_\alpha + 1) K}{m_\alpha n^{\text{ref}}} \ln \left(1 + \frac{p - p^{\text{ref}}}{K} \right) \right) = g_\alpha^{\text{ref}} + \frac{(\kappa_\alpha + 1)}{m_\alpha n^{\text{ref}}} (p - p^{\text{ref}}).
\end{equation}
%--------------------------------
Under incompressibility, the specific Gibbs energy depends linearly on pressure, with solvation effects captured via the solvation number \( \kappa_\alpha \).
The total number density becomes constant, i.e. $n = n^{\text{ref}}$, simplifying the model by reducing variables and focusing on electrochemical and solvation effects.

%%%%%%%%%%%%%%%%%%%%%%%%%%%%%%%%%%%%%%%%%%%%%%%%%%%%%%%%%%
\subsubsection{Dimensionless quantities}\label{Sec:2.5.3}

To facilitate analysis and numerical simulation, we introduce dimensionless variables by scaling the governing equations with appropriate reference quantities. We use $n^{\text{ref}}$, $p^{\text{ref}}$, $L^{\text{ref}}$, and the thermal voltage $kT_0/e_0$ as reference scales. The dimensionless variables (reusing the same symbols for brevity) are defined as follows~\cite{dreyer2013overcoming}:
%--------------------------------
\begin{equation}\label{eq:dimless_vars}
\begin{aligned}
 n &\rightarrow n^{\text{ref}} n, \quad p \rightarrow p^{\text{ref}} p, \quad z_\alpha \rightarrow e_0 z_\alpha, \quad K \rightarrow p^{\text{ref}} K, \quad g_\alpha \rightarrow \frac{kT_0}{m_\alpha} g_\alpha, \\
 \mu_{\alpha} &\rightarrow\frac{kT_{0}}{m_{\alpha}}\mu_{\alpha}, \quad n^F \rightarrow n^{\text{ref}} e_0 n^F, \quad \varphi \to \frac{kT_0}{e_0}\varphi, \quad x \rightarrow L^\text{ref} x.
\end{aligned}
\end{equation}
%--------------------------------

\subsection{Summary of the model}

To summarize, we consider a steady-state, dimensionless formulation on one- and two-dimensional domains with multiple ionic species. Using the dimensionless variables from~\eqref{eq:dimless_vars} in~\eqref{Eq:15} yields the final PDE system:
%==================================
\begin{equation}
	\begin{aligned}
		\label{Eq:23}
		\text{div} \, (\boldsymbol{J}_{\alpha}) & =0, \quad  \alpha\in\{1,\dots,N-1\}, \\
		a^2 \nabla p & =-n^{F}\nabla \varphi, \\
		\lambda^2 \, \Delta\varphi & =- \, n^{F},
	\end{aligned}
\end{equation}
%===================================
in which the two dimensionless parameters $\lambda^2$ and $a^2$ are defined as
%===================================
\begin{equation}
	\label{Eq:24}
	\lambda^2 = \frac{k T_0 \varepsilon_0 (1 + \chi)}{e_0^2 n^{\text{ref}} (L^{\text{ref}})^2}, \quad a^2 = \frac{p^{\text{ref}}}{n^{\text{ref}} k T_0}.
\end{equation}
%===================================
Here, $\lambda^2$ corresponds to the scaled Debye length $\lambda_{D}^{2}=\lambda^{2}(L^{\text{ref}})^{2}=kT_{0}\varepsilon_{0}(1+\chi)/(e_{0}^{2}n^{\text{ref}})$, characterizing the thickness of the electric double layer near charged surfaces, while $a^2$ is a pressure scaling factor, rescaling the pressure field without influencing other physical quantities. Furthermore, the dimensionless diffusion flux is expressed as
%===================================
\begin{equation}
	\label{Eq:25}
	\boldsymbol{J}_\alpha = \nabla(\mu_\alpha - \mu_N + z_\alpha \varphi),
\end{equation}
%===================================
since the \(M_{\alpha}\) from~\eqref{Eq:16} vanish with~\eqref{Eq:23}, while the chemical potential given by
%===================================
\begin{equation}
	\label{Eq:26}
	\mu_\alpha = g_\alpha + \ln(y_\alpha).
\end{equation}
%===================================
The specific Gibbs energy $g_\alpha$ in~\eqref{Eq:26} depends on the compressibility assumption (see Sec.~\ref{Sec:2.5.2}):
%==================================
\begin{equation}
	\begin{aligned}
		\label{Eq:27}
		% \text{Compressible case:} & \quad
		% g_\alpha = g_\alpha^{\text{ref}} + (\kappa_\alpha + 1) a^2 K \ln\left(1 + \frac{1}{K}(p - 1) \right), \\
		% \text{Incompressible case:} & \quad
		% g_\alpha = g_\alpha^{\text{ref}} + (\kappa_\alpha + 1) a^2 (p - 1).
		% \\
		g_\alpha &= \begin{cases}
			g_\alpha^{\text{ref}} + (\kappa_\alpha + 1) a^2 K \ln\left(1 + \frac{1}{K}(p - 1) \right) & \text{for the compressible case}, \\
			g_\alpha^{\text{ref}} + (\kappa_\alpha + 1) a^2 (p - 1) & \text{for the incompressible case},
		\end{cases}
	\end{aligned}
\end{equation}
%===================================
%===================================
while the dimensionless free charge density is given by
\begin{equation}
	\label{Eq:28}
	n^F = n \sum_{\alpha=1}^N z_\alpha y_\alpha.
\end{equation}
%===================================

%%%%%%%%%%%%%%%%%%%%%%%%%%%%%%%%%%%%%%%%%%%%%%%%%%%%%%%%%%%
%%%%%%%%%%%%%%%%%%%%%%%%%%%%%%%%%%%%%%%%%%%%%%%%%%%%%%%%%%
\section{Numerical procedure}\label{Sec:3}

This section briefly describes the FEM used to numerically solve the thermodynamically consistent electrolyte model. FEM efficiently handles coupled, nonlinear PDEs over complex geometries with heterogeneous boundary conditions. The numerical approach involves domain discretization, selecting suitable finite elements, and applying boundary conditions. These steps underpin the weak formulation and solver implementation within the FEniCS framework.

%%%%%%%%%%%%%%%%%%%%%%%%%%%%%%%%%%%%%%%%%%%%%%%%%%%%%%%%%%
\subsection{Implementation using FEniCS}\label{Sec:3.1}

The coupled nonlinear PDEs (\ref{Eq:23}) are numerically solved using FEM implemented within the open-source computing platform FEniCS~\cite{BarattaEtal2023,logg2012finite}. FEniCS is a robust software suite designed specifically to automate and simplify the solutions of PDEs. It comprises core components, including DOLFINx, FFCx, Basix, and UFL, and provides intuitive interfaces in Python and C++.
The FEniCS framework streamlines finite element solver development by enabling users to easily specify the variational formulation of PDE systems, function spaces, element types, boundary conditions, and computational meshes. Once defined, FEniCS automatically handles underlying operations such as element-wise assembly, matrix formation, and solving the resulting algebraic system. This automated approach significantly reduces manual coding efforts, minimizes the potential for errors, enhances reproducibility, and accelerates computational modeling workflows.

%%%%%%%%%%%%%%%%%%%%%%%%%%%%%%%%%%%%%%%%%%%%%%%%%%%%%%%%%%
\subsection{Variational formulation and boundary implementation}\label{Sec:3.2}

The FEM implementation is based on a mixed variational formulation
of the dimensionless governing equations. The approach relies on standard
Sobolev space theory for elliptic equations and weak formulations
derived via integration by parts, suitable for implementation in the
FEM framework. We denote the spatial domain by $\Omega\subset\mathbb{R}^{n}$
and let $H^{m}(\Omega)$ denote the Sobolev space of order $m$, defined
as
\begin{equation}
H^{m}(\Omega)=\left\{ u\in L^{2}(\Omega)\ \middle|\ D^{\alpha}u\in L^{2}(\Omega),\ |\alpha|\leq m\right\}, \label{Eq:30}
\end{equation}
where $D^{\alpha}u$ denotes weak derivatives of $u$, and $L^{2}(\Omega)$
is the space of square-integrable functions.

The primary unknowns are the electric potential $\varphi$, pressure
$p$, and atomic fractions $y_{\alpha}$ for $\alpha\in\{1,\dots,N-1\}$
for each of which boundary conditions can be of Dirichlet- and Neumann-type.
We assume that the boundary $\partial\Omega$ is divided into disjoint
Dirichlet and Neumann parts individually for the different variables,
i.e.,
\begin{equation}
\partial\Omega^{(w)}=\Gamma_{D}^{(w)}\cup\Gamma_{N}^{(w)},\quad\Gamma_{D}^{(w)}\cap\Gamma_{N}^{(w)}=\emptyset,\quad w\in\{\varphi,p,y_{\alpha}\}.\label{Eq:29}
\end{equation}
The Neumann boundary conditions are given as
\begin{equation}
\begin{aligned}\label{Eq:34a}\nabla\varphi\cdot\boldsymbol{n} & =h_{\varphi}\quad\,\,\bm{x}\in\Gamma_{N}^{(\varphi)},\\
\left(\nabla p+\frac{1}{a^{2}}n^{F}\nabla\varphi\right)\cdot\boldsymbol{n} & =h_{p}\quad\,\,\bm{x}\in\Gamma_{N}^{(p)},\\
J_{\alpha}\cdot\boldsymbol{n} & =h_{y_{\alpha}}\quad\bm{x}\in\Gamma_{N}^{(y_{\alpha})},\quad\alpha\in\{1,\dots,N-1\},
\end{aligned}
\end{equation}
where $\boldsymbol{n}$ is the outward-pointing unit normal vector and
Dirichlet conditions take the form
\begin{equation}
\begin{aligned}\label{Eq:dirichlet}
	\varphi & =g_{\varphi}\quad\,\,\bm{x}\in\Gamma_{D}^{(\varphi)},\\
	p & =g_{p}\quad\,\,\bm{x}\in\Gamma_{D}^{(p)},\\
	y_{\alpha} & =g_{y_{\alpha}}\quad\bm{x}\in\Gamma_{D}^{(y_{\alpha})},\quad\alpha\in\{1,\dots,N-1\},
\end{aligned}
\end{equation}
where $g_{\varphi},g_{p},g_{y_{\alpha}}$ and $h_{\varphi},h_{p},h_{y_{\alpha}}$are given,  process-depending, functions. Throughout this paper we will use homogeneous
Neumann conditions for pressure, i.e., $h_{p}=0$ and $\partial\Omega^{(p)}=\Gamma_{N}^{(p)}$.
Most examples will be driven by a prescription of different potential
values, $g_{\varphi}\ne0$ at specific parts of the boundary while
setting the mass flux to zero on all of $\partial\Omega$, that is,
$h_{y_{\alpha}}=0$. Occasionally, Dirichlet conditions for the mol fractions are also used.

We define the trial and test function spaces as:
\begin{equation}
\begin{aligned}
	V^{(w)} & =\left\{ u\in H^{1}(\Omega)\ \middle|\ u=g_{w}\text{ on }\Gamma_{D}^{(w)}\right\} ,\quad\quad V_{0}={\left\{ u\in H^{1}(\Omega)\ \middle|\ u=0\text{ on }\partial\Omega\right\}},
\end{aligned}
\label{Eq:32}
\end{equation}
and construct the variational problem:

Find $(\varphi,p,y_{\alpha})\in V^{(\varphi)}\times V^{(p)}\times\prod_{\alpha=1}^{N-1}V^{(y_{\alpha})}$,
such that for all test functions $(v_{\varphi},v_{p},v_{y_{\alpha}})\in[V_{0}]^{N+1}$,
\begin{equation}
\begin{aligned}\label{Eq:33}0= & \int_{\Omega}\nabla\varphi\cdot\nabla v_{\varphi}\,\text{d}\Omega-\int_{\Gamma_{N}^{(\varphi)}}h_{\varphi}v_{\varphi}\,\text{d}s-\int_{\Omega}\frac{1}{\lambda^{2}}n^{F}v_{\varphi}\,\text{d}\Omega+\int_{\Omega}\left(\nabla p+\frac{1}{a{^{2}}}n^{F}\nabla\varphi\right)\cdot\nabla v_{p}\,\text{d}\Omega\\
 & -\int_{\Gamma_{N}^{(p)}}h_{p}v_{p}\,\text{d}s+\sum_{\alpha=1}^{N-1}\Bigg(\int_{\Omega}\boldsymbol{J}_{\alpha}\cdot\nabla v_{y_{\alpha}}\,\text{d}\Omega-\int_{\Gamma_{N}^{(y_{\alpha})}}h_{y_{\alpha}}v_{y_{\alpha}}\,\text{d}s\Bigg).
\end{aligned}
\end{equation}
Here, $J_{\alpha}$ denotes the diffusion flux for species $\alpha$ from~\eqref{Eq:25},
which still depends nonlinearly on $\varphi$, $p$, and $y_{\alpha}$ by~\eqref{Eq:26} and~\eqref{Eq:27}.
The space charge density $n^{F}$ is a function of the atomic fractions
$y_{\alpha}$ by~\eqref{Eq:28}. In this work,
standard first-order Lagrange finite elements are used to approximate
all variables.

%%%%%%%%%%%%%%%%%%%%%%%%%%%%%%%%%%%%%%%%%%%%%%%%%5

%%%%%%%%%%%%%%%%%%%%%%%%%%%%%%%%%%%%%%%%%%%%%%%%%%%%%%%%%%
\subsection{Numerical treatment of the nonlinearity}\label{Sec:3.3}

The governing equations introduced in Section~\ref{Sec:2} contain strong nonlinearities arising from the exponential dependence of the chemical potentials on mole fractions and pressure, as well as from the coupling of electrostatic potential and ion transport. To solve the resulting nonlinear variational system, we employ a Newton-Raphson method provided by the FEniCS finite element framework~\cite{BarattaEtal2023, logg2012finite}. The Newton method iteratively linearizes the nonlinear system using the Jacobian matrix and updates the solution via:
%-------------------------------
%-------------------------------
\begin{equation}
\label{Eq:34b}
\boldsymbol{x}^{(k+1)} = \boldsymbol{x}^{(k)} - \gamma \left[ J_{\boldsymbol{F}}(\boldsymbol{x}^{(k)}) \right]^{-1} \boldsymbol{F}(\boldsymbol{x}^{(k)}),
\end{equation}
%-------------------------------
%-------------------------------
where $\boldsymbol{x}^{(k)}$ denotes the solution vector at the $k$-th iteration, $\boldsymbol{F}$ is the residual vector representing the weak form of the governing equations, $J_{\boldsymbol{F}}$ is the Jacobian of $\boldsymbol{F}$, and $\gamma \in (0,1]$ is a relaxation parameter introduced to improve convergence stability.  % chktex 9
In practice, setting $\gamma = 1$ often leads to divergence due to the sensitivity of the nonlinear system, particularly when sharp gradients occur near charged interfaces. As a remedy, we use a fixed damping strategy with a manually tuned relaxation parameter $\gamma \ll 1$, chosen based on empirical testing across different benchmark problems. This approach maintains robustness while minimizing unnecessary complexity.

%%%%%%%%%%%%%%%%%%%%%%%%%%%%%%%%%%%%%%%%%%%%%%%%%%%%%%%%%%%
%%%%%%%%%%%%%%%%%%%%%%%%%%%%%%%%%%%%%%%%%%%%%%%%%%%%%%%%%%
\section{Convergence study and validation}\label{Sec:4}

To verify the accuracy and reliability of the developed computational model, we perform validation tests against the established benchmark problems and analytical solutions available in the literature. Key aspects assessed include solution accuracy, convergence behavior, and consistency with thermodynamic constraints. The solver's predictions are systematically compared with reference results to demonstrate its capability for accurately modeling electrolyte systems under various electrochemical conditions and physical parameters.
In the subsequent simulations, we use the following meaningful, physical values:
%------------------------------
%-------------------------------
\begin{equation*}
\begin{aligned}
	T & =293.75 \, \text{K}, \quad  n^{\text{ref}} = 55 \cdot N_{A} \cdot10^{-3} \, m^{-3},  \quad p^{\text{ref}} = 1.01325\cdot10^{5} \, \text{Pa}, \\
	L^{\text{ref}} & = 20\cdot10^{-9} \, \text{m}, \quad \chi = 80, \quad
	\lambda^{2}  =8.553\cdot10^{-6}, \quad  a^{2} =7.5412\cdot10^{-4}.
\end{aligned}
\end{equation*}
%-------------------------------
%------------------------------
The values for $T$, $n^{\text{ref}}$, $p^{\text{ref}}$, and $\chi$ are taken from~\cite{dreyer2014mixture}. Since the total length of the domain used was not specified, $L^{\text{ref}}$ is set to $20 \times 10^{-9} \, \text{m}$, based on the value provided in~\cite{Fuhrmann_LiquidElectrolytes_GitHub}. In the following sections, we use the notation $\kappa = \kappa_{\alpha}$ if the solvation number is constant for all charged constituents $(\alpha \in \{1,\ldots,N-1\})$. If not otherwise stated, we set $\kappa = 0$ and assume incompressibility (see Sec.~\ref{Sec:2.5.2}) for all simulations.

%$$$$$$$$$$$$$$$$$$$$$$$$$$$$$$$$$$$$$$$$$$$
\begin{figure}
	\centering
	\includegraphics[width=.9\textwidth]{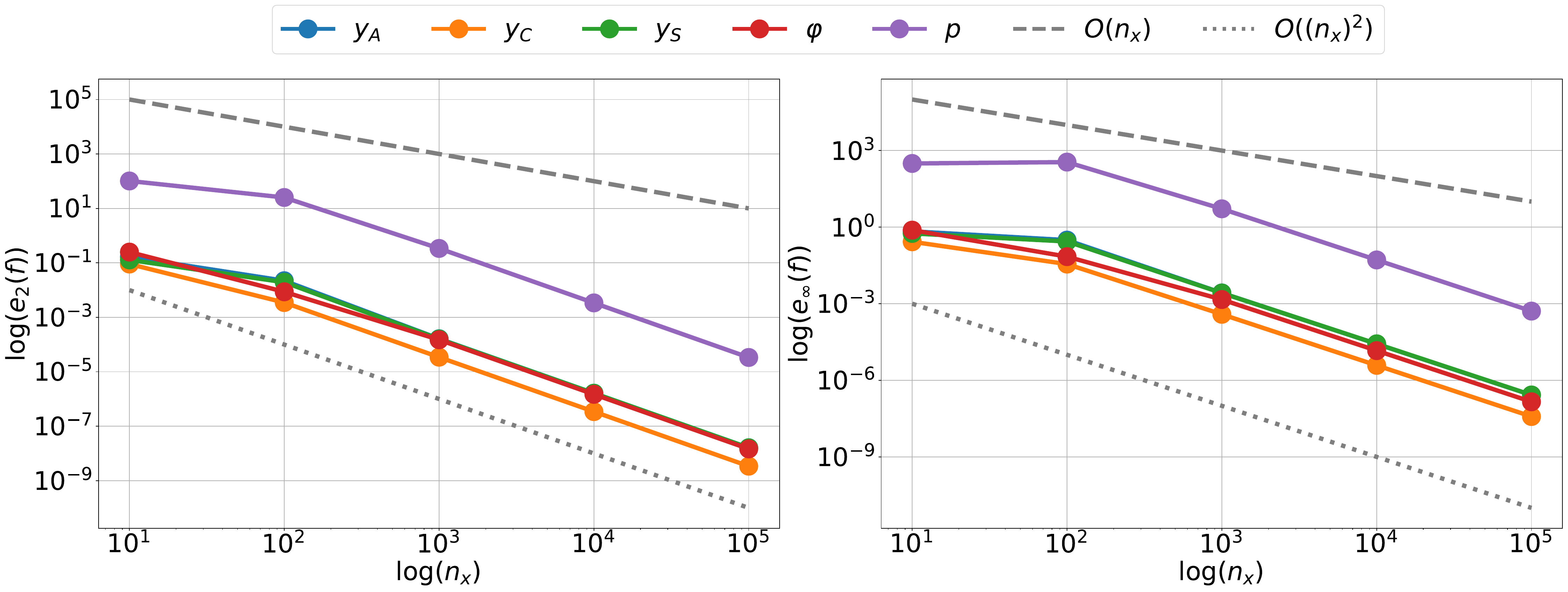}
	\caption{Convergence behavior of the finite element solver for a one-dimensional incompressible ternary electrolyte system. The plot shows the $e_{2}-$ and $e_{\infty}-$error norms as functions of mesh resolution (number of elements, $n_x$) on a logarithmic scale.}\label{Fig:2}
\end{figure}
%$$$$$$$$$$$$$$$$$$$$$$$$$$$$$$$$$$$$$$$$$$$

%%%%%%%%%%%%%%%%%%%%%%%%%%%%%%%%%%%%%%%%%%%%%%%%%%%%%%%%%%
\subsection{Convergence study}\label{Sec:4.1}

A convergence study is performed using a one-dimensional incompressible ternary electrolyte system to assess the numerical accuracy of the proposed finite element solver. Due to the lack of an analytical solution, convergence is evaluated by comparing numerical results from coarser grids against a reference solution computed on a highly refined mesh with $n_x=1,000,000$  equally sized elements.

The convergence behavior is quantified using two standard error norms:
%%%%%%%%%%%%%%%%%%%%%%%%%%%%%%%%%%%%%%%%%%%%%%%%
\begin{equation}
	e_2(f_h) = \left(\int_\Omega (f_{\text{fine}} - f_h)^2 \,\text{d}\Omega \right)^{1/2}, \quad
	e_\infty(f_h) = \max_{x \in \eta} |f_{\text{fine}}(x) - f_h(x)|,
\end{equation}
%%%%%%%%%%%%%%%%%
where $f_{\text{fine}}$ is the solution on the finest mesh, $f_{h}$ is the solution on a coarser mesh with mesh size $h$, and $\eta$ is the set of all mesh nodes.

The test problem considered a ternary electrolyte comprising negatively charged anions ($y_{A}, \,\, z_A=-1$), positively charged cations ($y_{c}, \,\,z_C=1$), and a neutral solvent ($y_{s}, \,\, z_s=0$). The applied potential difference was $\delta\varphi=6$, with vanishing pressure and atomic fractions ($y_{\alpha}=1/3$) specified at the right boundary of the computational domain. The dimensionless parameters were set to $\lambda{{}^2}=8.553\cdot10^{-6}$ and $a^{2}=7.5412\cdot10^{-4}$.

Figure~\ref{Fig:2} illustrates the error measured in the $e_{2}$- and $e_{\infty}$-norms as a function of the number of elements ($n_x$) on a logarithmic scale. Both norms exhibit second-order convergence, which is the behavior expected when using first-order elements. Although the FEniCS framework supports higher-order discretization schemes (and extension to such elements is straightforward), we restrict our analysis to the first-order elements specified because they provide sufficient accuracy for the present study. The convergence of both error norms toward zero confirms the solver's consistency and its ability to approximate the reference solution across the entire domain with increasing resolution.

%%%%%%%%%%%%%%%%%%%%%%%%%%%%%%%%%%%%%%%%%%%%%%%%%%%%%%%%%%
\subsection{Validation study}\label{Sec:4.2}

\begin{figure}[t]
	\centering
	\includegraphics[width=0.9\textwidth]{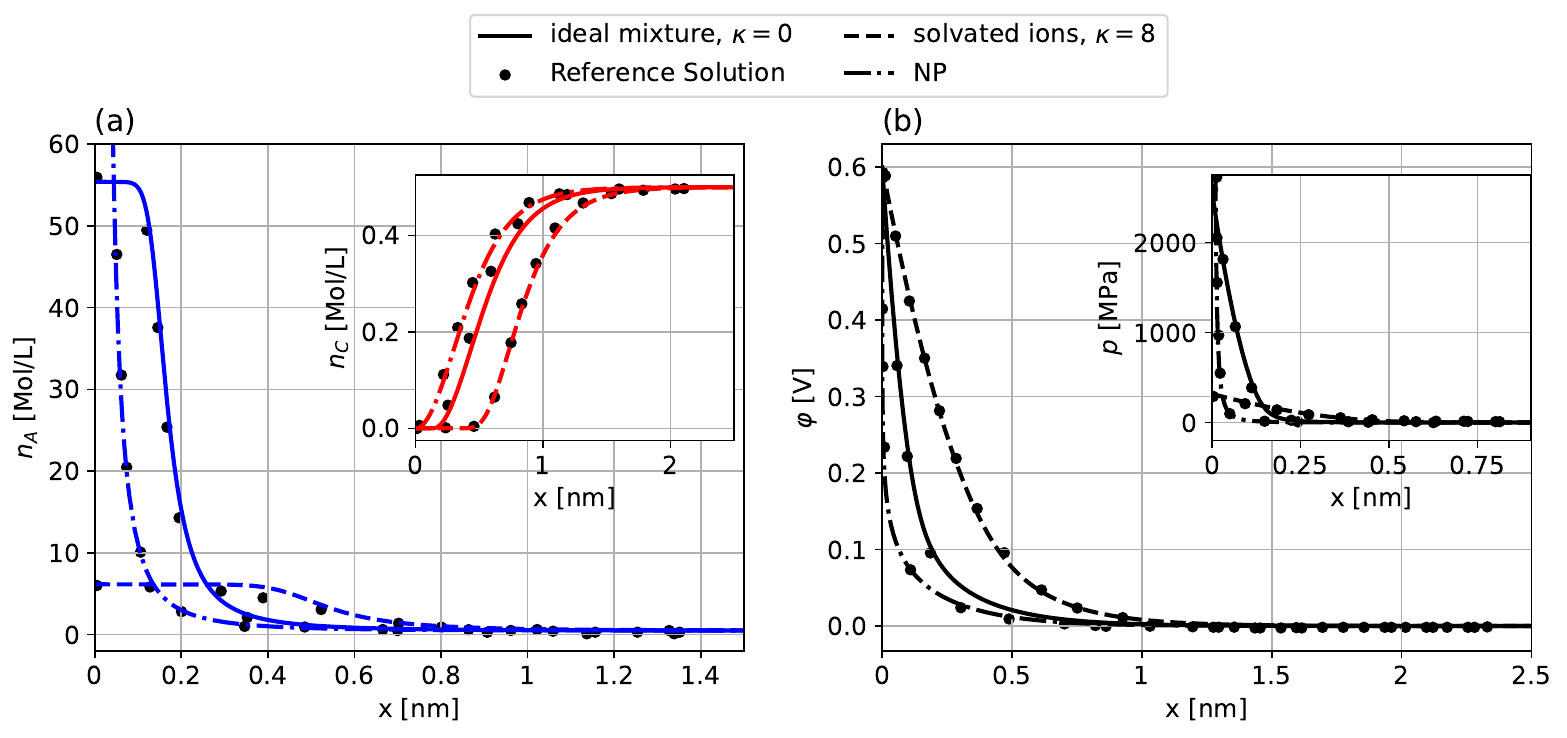}
	\caption{Validation study comparing the thermodynamically consistent model with the classical NP model in a one-dimensional ternary electrolyte system. (a) Number density profiles of anions ($n_A$) and cations ($n_C$) for three model variants: ideal mixture ($\kappa = 0$), solvated ions ($\kappa = 8$), and the classical NP model in Poisson--Boltzmann formulation. (b) Electrostatic potential ($\varphi$) and pressure ($p$) distributions for the same configurations. Dots indicate the reference solution from~\cite{dreyer2018bulk}. The results demonstrate close agreement with the reference and highlight the divergence behavior in the NP model, emphasizing the improved physical consistency of the extended model.}\label{Fig:3}  % chktex 12
\end{figure}

To validate the numerical implementation of the thermodynamically consistent electrolyte model, a benchmark simulation from Dreyer et al.~\cite{dreyer2018bulk} is reproduced. The test scenario involves a one-dimensional, incompressible ternary electrolyte system subjected to an applied dimensionless potential difference of 0.6V for an electrolyte with a molarity of 0.5mol/L. As the original reference does not provide complete parameter specifications, particularly for the scaled Debye length $\lambda$ and pressure scaling factor $a$, these quantities are set to $\lambda^2=2.4 \times 10^{-6}$, and $a^2=7.5 \times 10^{-4}$. Since these parameters influence the spatial scaling, a degree of uncertainty remains in the horizontal alignment of the reproduced results, which has been used to obtain the best fit. Figure~\ref{Fig:3} illustrates a comparison among three different model variants: the solid line represents the thermodynamically consistent model incorporating solvation effects with a solvation number of $\kappa = 8$; the dash-dotted line corresponds to the same model assuming an ideal mixture without solvation ($\kappa = 0$); and the dotted line displays the solution obtained from the classical NP model in its Poisson--Boltzmann formulation. The left-hand plots show the number densities of anions and cations, while the right-hand plots depict the electric potential and pressure distributions. The overall agreement with the reference solution confirms the accuracy of the implementation, accounting for the uncertainty introduced by unknown scaling parameters. Importantly, the classical NP model exhibits nonphysical behavior, such as the divergence of anion concentration near the charged interface, highlighting its inadequacy in capturing non-ideal effects. In contrast, the thermodynamically consistent model accurately reflects the influence of solvation and electrochemical coupling, underscoring the necessity of incorporating such effects for realistic electrolyte modeling.

%$$$$$$$$$$$$$$$$$$$$$$$$$$$$$$$$$$$$$$$$$$$
%$$$$$$$$$$$$$$$$$$$$$$$$$$$$$$$$$$$$$$$$$$$

%%%%%%%%%%%%%%%%%%%%%%%%%%%%%%%%%%%%%%%%%%%%%%%%%%%%%%%%%%%
%%%%%%%%%%%%%%%%%%%%%%%%%%%%%%%%%%%%%%%%%%%%%%%%%%%%%%%%%%
\section{Results and discussions}\label{Sec:5}
We are now prepared to present numerical simulations performed using the implemented solver. For reproducibility, the source code and all metadata are available at~\cite{Habscheid_2025}.

\subsection{One-dimensional ternary electrolyte}\label{Sec:5.1}

A ternary electrolyte consists of three constituents ($N = 3$): negatively charged anions ($A$, $\alpha = 1$), positively charged cations ($C$, $\alpha = 2$), and a neutral solvent ($S$, $\alpha = 3$). In this configuration, only the anions and cations contribute to the space charge, while the solvent is electrically neutral with charge number $z_N = 0$. Here, and in the remaining sections, we identify component $N$ with the neutral solvent $S$.

The total space charge density of the ternary electrolyte is given by
%%%%%%%%%%%%%%%%%%%%%%%%%%%%%%%%%%%%%
%%%%%%%%%%%%%%%%%%%%%%%%%%%%%%%%%%%%%%%%%%%%%%%%
\begin{equation}
	n_{\text{ternary}}^F = \sum_{\alpha=1}^{3} z_\alpha e_0 n_\alpha  = e_0(z_A n_A + z_C n_C + z_N n_N)
	= e_0(z_A n_A + z_C n_C),
\end{equation}
%%%%%%%%%%%%%%%%%
where $e_0$ is the elementary charge, $z_\alpha$ are the charge numbers, and $n_\alpha$ the number densities of each species.

This formulation simplifies the general multi-component electrolyte system and serves as a physically meaningful model for various applications such as batteries and desalination. The neutral solvent, while not affecting the space charge, is essential for capturing solvation and non-ideal transport effects, which are integrated into the thermodynamically consistent model.

For the numerical experiments, the ionic charges are set as $z_A = -1$, $z_C = 1$, and $z_N = 0$. The following Dirichlet boundary conditions are imposed on the right boundary of the domain:
%---------------------------------------
\begin{equation*}
	\varphi^R = 0.0, \quad p^R = 0.0, \quad y_A^R = y_C^R = \frac{1}{3}.
\end{equation*}
%---------------------------------------
The computational domain spans $x = [0, 1]$, with the inner region of the electrolyte---representing the bulk---located on the right. The left portion near the boundary layer experiences significant variations, while the remainder of the domain remains in quasi-equilibrium. To focus on the region of physical interest, only the subdomain $x_{\text{vis}} = [0, 0.025]$ is visualized, as the solution remains nearly constant beyond this range. This localized behavior highlights the complexity and nonlinearity of electrochemical interactions near interfaces, which are effectively captured by the present model.

%$$$$$$$$$$$$$$$$$$$$$$$$$$$$$$$$$$$$$$$$$$$
\begin{figure}
	\centering
	\includegraphics[width=.9\textwidth]{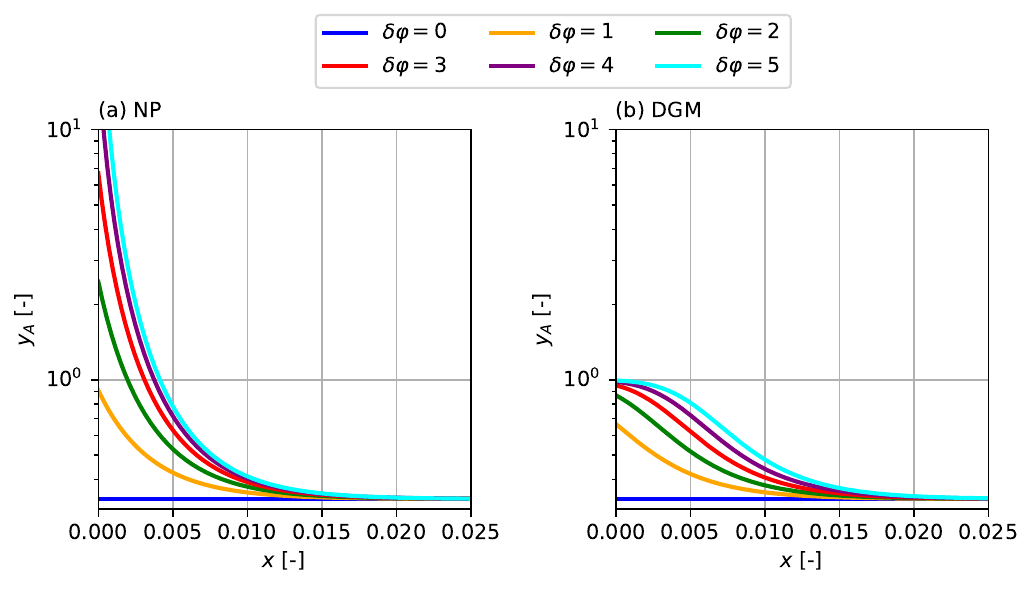}
	\caption{Comparison between (a) classical NP model, and (b) DGM model: solution profiles of anion atomic fraction at different potential differences.}\label{Fig:4}
\end{figure}
%$$$$$$$$$$$$$$$$$$$$$$$$$$$$$$$$$$$$$$$$$$$

%%%%%%%%%%%%%%%%%%%%%%%%%%%%%%%%%%%%%%%%%%%%%%%%%%%%%%%%%%
\subsubsection{Comparison to the classical NP model}\label{Sec:5.1.1}

The thermodynamically consistent electrolyte model utilized in this paper (hereafter referred to as the DGM model), following~\cite{dreyer2013overcoming}, addresses key limitations of the classical Nernst--Planck (NP) model and its Poisson--Boltzmann formulation. One central distinction is the proper coupling of diffusion fluxes. In the DGM model, only $N-1$ fluxes are treated as independent, while the flux of the remaining species is obtained from the constraint
\begin{equation}
	\sum_{\alpha=1}^{N} \boldsymbol{J}_\alpha = 0,
\end{equation}
which ensures mass conservation and consistency with the barycentric frame. Consequently, the model formulation leads to $N - 1$ coupled diffusion equations instead of $N$ uncoupled ones, as seen in the classical NP framework.

Furthermore, the DGM model includes additional physical effects such as pressure coupling and the influence of the chemical potential of the neutral solvent. These aspects are entirely absent in the classical NP model. As a result, the NP model fails to capture saturation behavior in ionic concentrations and often produces nonphysical solutions---particularly in regions with strong electrostatic fields or near charged boundaries. In equilibrium, the atomic fractions predicted by the NP model exhibit exponential dependence on the electrostatic potential, which can cause them to exceed the physically admissible range of $[0,1]$.

%$$$$$$$$$$$$$$$$$$$$$$$$$$$$$$$$$$$$$$$$$$$
\begin{figure}
	\centering
	\includegraphics[width=.9\textwidth]{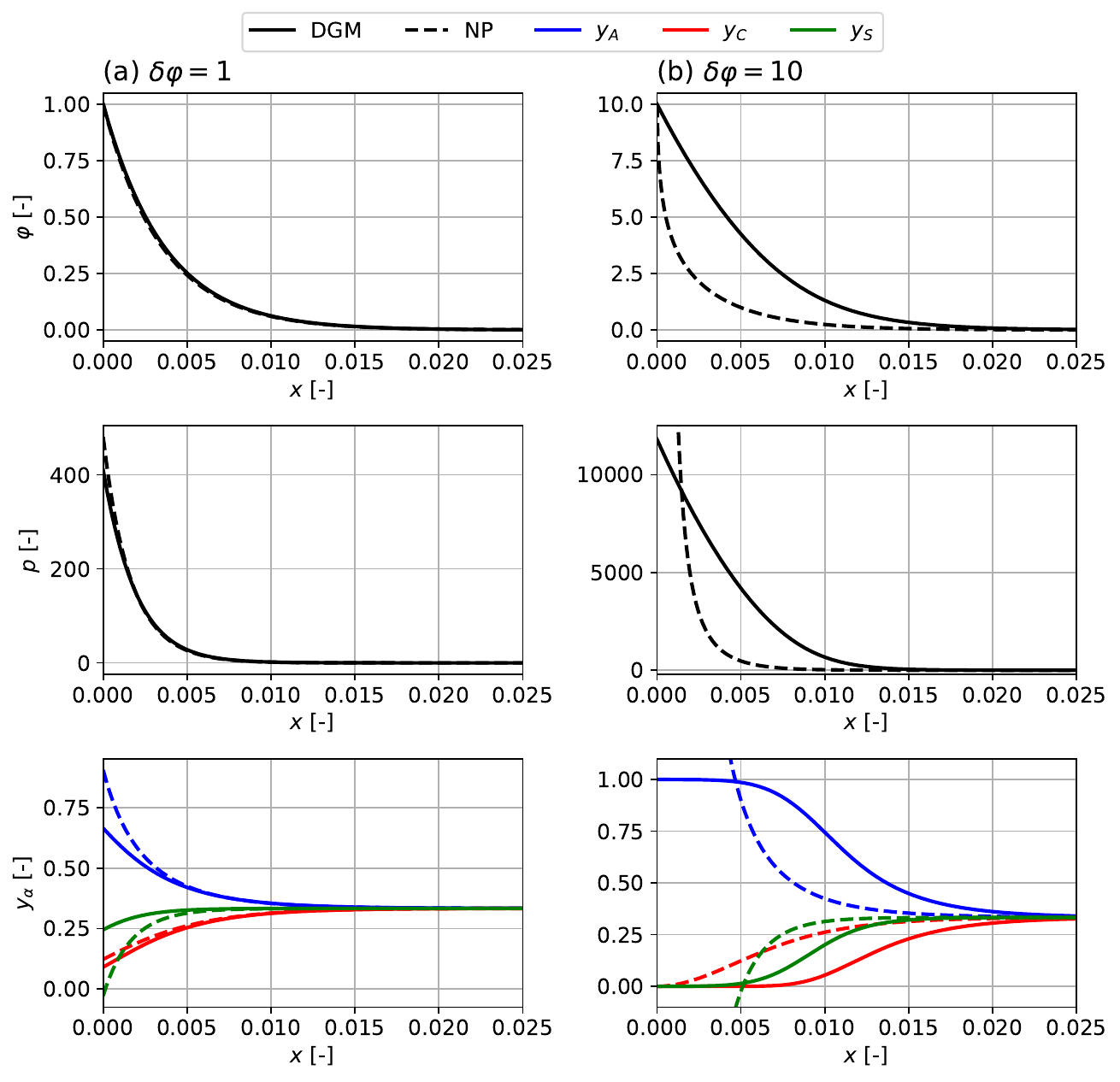}
	\caption{Comparison between DGM and the classical NP model: profiles of electric potential ($\varphi$), pressure ($p$) and atomic fraction ($y_{\alpha}$) at different electric potentials (a) $\delta \varphi=1$, and (b) $\delta \varphi = 10$.}\label{Fig:5}
\end{figure}
%$$$$$$$$$$$$$$$$$$$$$$$$$$$$$$$$$$$$$$$$$$$

Figure~\ref{Fig:4} compares the anion atomic fraction profiles computed using the classical NP and thermodynamically consistent DGM models under varying applied potential differences. It clearly illustrates that the NP model yields unrealistic concentrations that violate physical bounds, especially near interfaces. In contrast, the DGM model maintains the solution within admissible limits, preventing divergence and ensuring consistency with thermodynamic constraints. The reason for this behavior lies in the DGM model's incorporation of the constraint $\sum_{\alpha=1}^N y_\alpha = 1$ (see also Section~\ref{Sec:2.4}) and the logarithmic dependence of the chemical potential on the atomic fractions. As the applied potential increases, the DGM model enforces saturation: the anion fraction $y_A$ approaches unity near the electrode, while the other fractions decrease toward zero, but none can exceed the physical bounds $[0,1]$. This is a direct consequence of the entropy term in the chemical potential (see~\eqref{Eq:26}), which becomes singular as any $y_\alpha \to 0$ or $y_\alpha \to 1$. In contrast, the classical NP model lacks this coupling and constraint, allowing concentrations to diverge exponentially with increasing potential, which is nonphysical.

Additionally, Figure~\ref{Fig:5} presents a broader comparison of electric potential ($\varphi$), pressure ($p$), and atomic fraction ($y_\alpha$) profiles at applied voltages $\delta \varphi = 1$ and $\delta \varphi = 10$. As the applied voltage increases, the classical NP model produces diverging ion concentrations, especially for anions and the neutral solvent, while the DGM model exhibits saturation behavior, limiting atomic fractions to the physical range. These discrepancies become more pronounced under strong electric fields, where the DGM model better reflects the physical blocking of ionic species.

Overall, the comparison underscores the deficiencies of the classical NP model and highlights the importance of thermodynamic consistency in accurately modeling ion transport in concentrated and highly coupled systems. The DGM model not only ensures mass and charge conservation but also respects entropy production and nonlinear coupling among species, offering a more reliable framework for electrolyte simulations.

\begin{figure}
	\centering
	\includegraphics[width=0.5\linewidth]{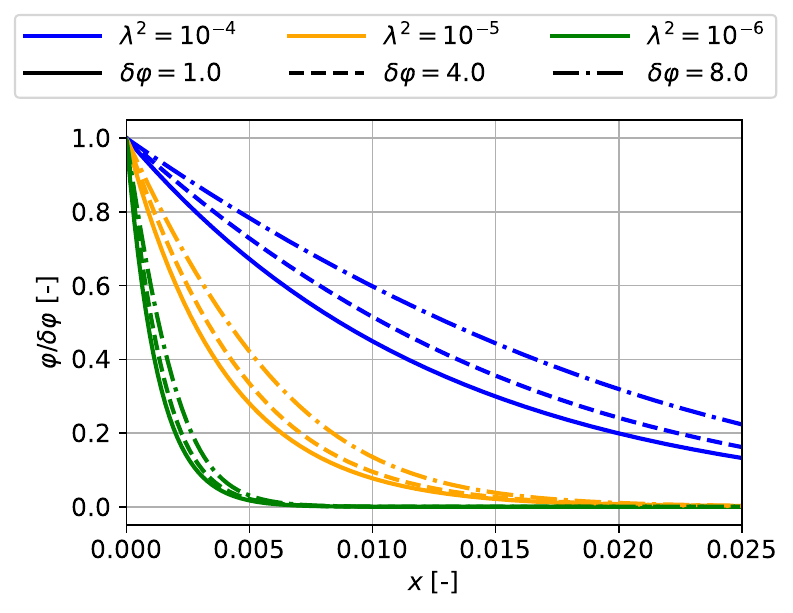}
	\caption{Combined effects of potential difference and scaled Debye length on the dynamics of three different electric potentials $\delta \varphi = 1.0, 4.0, 8.0$ with three different scaled Debye length $\lambda^{2}=10^{-4}, 10^{-5}, 10^{-6}$.}\label{Fig:6}
\end{figure}

Additionally, we note that while the ideal-mixing entropy captures steric saturation at high concentrations and strong fields, it does not account for explicit ionic diameters or nonlocal excluded-volume correlations. Consequently, the model does not capture packing-induced layering (oscillations) in the EDL structure, resulting in non-monotonic profiles in Figs.~\ref{Fig:4}--\ref{Fig:6} (and subsequent figures). To recover such effects, one may employ more sophisticated theories using more advanced nonlocal free energy models (see the discussion in Sec.~\ref{Sec:6}).

%================================
%================================
%================================

%%%%%%%%%%%%%%%%%%%%%%%%%%%%%%%%%%%%%%%%%%%%%%%%%%%%%%%%%%
\subsubsection{Role of various model parameters}\label{Sec:5.1.2}
We can now discuss the influence of the scaled Debye length, the electric potential, the solvation number, and the compressibility.

\paragraph{Combined effects of electric potential and scaled Debye length}
%%%%%%%%%%%%%%%%%%%%%%%%%%%%%%%%%%%%%%%%%%%%%%%%%%%%%%%%%%

Figure~\ref{Fig:6} systematically illustrates the coupled effects of the applied potential difference ($\delta \varphi$) and the scaled Debye length ($\lambda^2$) on the profile of the normalized electric potential $(\varphi/\delta\varphi)$ in a one-dimensional ternary electrolyte system. The different cases correspond to $\delta \varphi = 1.0$, $4.0$, and $8.0$, respectively. For each case, three different values of the scaled Debye length are considered: $\lambda^{2}=10^{-4}, 10^{-5}, 10^{-6}$.
The scaled Debye length, which characterizes the electrostatic screening effect within the electrolyte, significantly influences the shape of the normalized electric potential profile. As $\lambda^2$ increases, the potential decays more gradually, resulting in a broader and smoother transition from the boundary value to the bulk. Conversely, smaller values of $\lambda^2$ lead to sharper gradients near the interface, indicating a thinner electrical double layer and stronger localization of the potential drop.
In contrast, the applied potential difference primarily governs the range of $\varphi$. Indeed, without the normalization, the applied potential difference $\delta\varphi$ would act as a scaling parameter in the vertical direction, while $\lambda^2$ dictates the spatial decay and curvature of the profile.
The fact that the normalized curves do not collapse onto one another clearly indicates that, in the thermodynamically consistent model, the electrostatic potential is non-linearly coupled to other fields---such as the pressure and chemical potential gradients.

%================================
%================================
\begin{figure}
	\centering
	\includegraphics[width=1.0\linewidth]{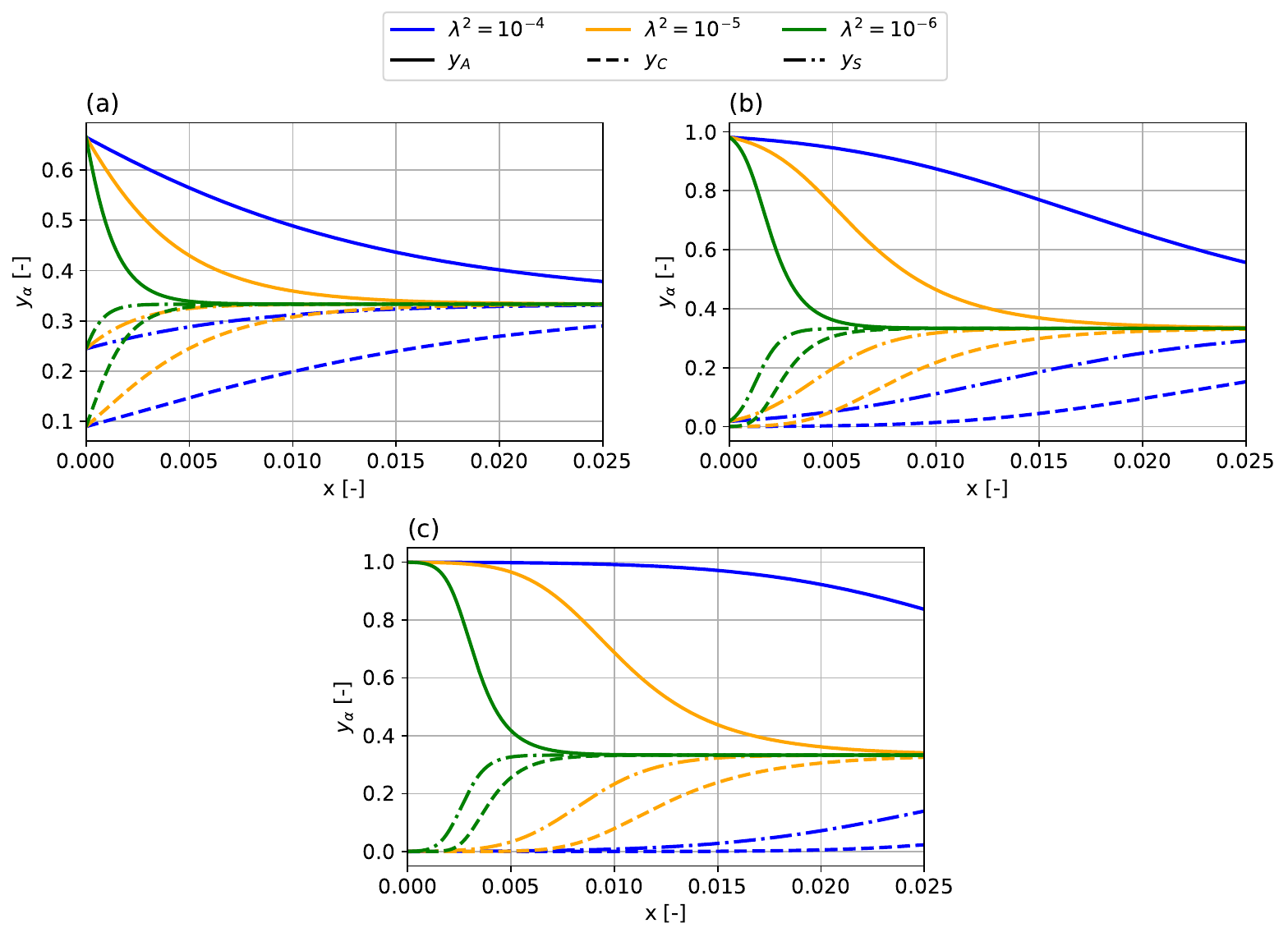}
	\caption{Combined effects of potential difference and scaled Debye length on the dynamics of atomic fractions: (a) $\delta \varphi = 1.0$. (c) $\delta \varphi = 4.0$, and (d) $\delta \varphi = 8.0$ with three different scaled Debye length $\lambda^{2}=10^{-4}, 10^{-5}, 10^{-6}$.}\label{Fig:7}  % chktex 12
\end{figure}
%================================

Figure~\ref{Fig:7} illustrates the combined effects of the scaled Debye length ($\lambda^2$) and the applied potential difference ($\delta \varphi$) on the spatial profiles of atomic fractions in a one-dimensional ternary electrolyte. The figure is subdivided into three panels corresponding to increasing values of the potential difference: (a) $\delta \varphi = 1.0$, (b) $\delta \varphi = 4.0$, and (c) $\delta \varphi = 8.0$. For each potential difference, simulations are carried out using three different values of the scaled Debye length: $\lambda^{2}=10^{-4}, 10^{-5}, 10^{-6}$. The line styles represent different species: solid lines for anions ($y_C$), dashed lines for cations ($y_A$), and dash-dotted lines for the neutral solvent ($y_S$).
The atomic fraction profiles reflect the formation and structure of the electrical double layer in response to electrostatic forces. As the applied potential difference increases, the deviation from the bulk (right boundary) values becomes more pronounced, particularly near the charged interface at the left boundary. A higher potential difference drives stronger ion separation, resulting in a higher accumulation of anions and a depletion of cations near the electrode for positive $\delta \varphi$. The reverse behavior would be observed for negative $\delta \varphi$ (not shown), where cations dominate near the negatively charged boundary. Smaller values of $\lambda^2$ correspond to thinner double layers with steep gradients, while larger values yield more gradual transitions and broader double layers. This behavior is rooted in the physical interpretation of the Debye length, which depends inversely on the reference number density and directly on temperature and dielectric susceptibility. Thus, reducing $\lambda^2$ effectively compresses the region over which electrostatic effects are significant. Notably, the solvent fraction ($y_S$) reaches its maximum at the electrode when no potential difference is applied, where the system remains in equilibrium with uniform concentrations. As $\delta \varphi$ increases, the solvent is displaced from the boundary layer to accommodate the accumulation of charged species, resulting in a decreasing $y_S$ near the interface. Despite the strong gradients induced by electrostatic forces, the atomic fractions remain constrained within the physically meaningful bounds of $[0, 1]$, demonstrating the robustness of the thermodynamically consistent model.

%================================
%================================
\begin{figure}
	\centering
	\includegraphics[width=1.0\linewidth]{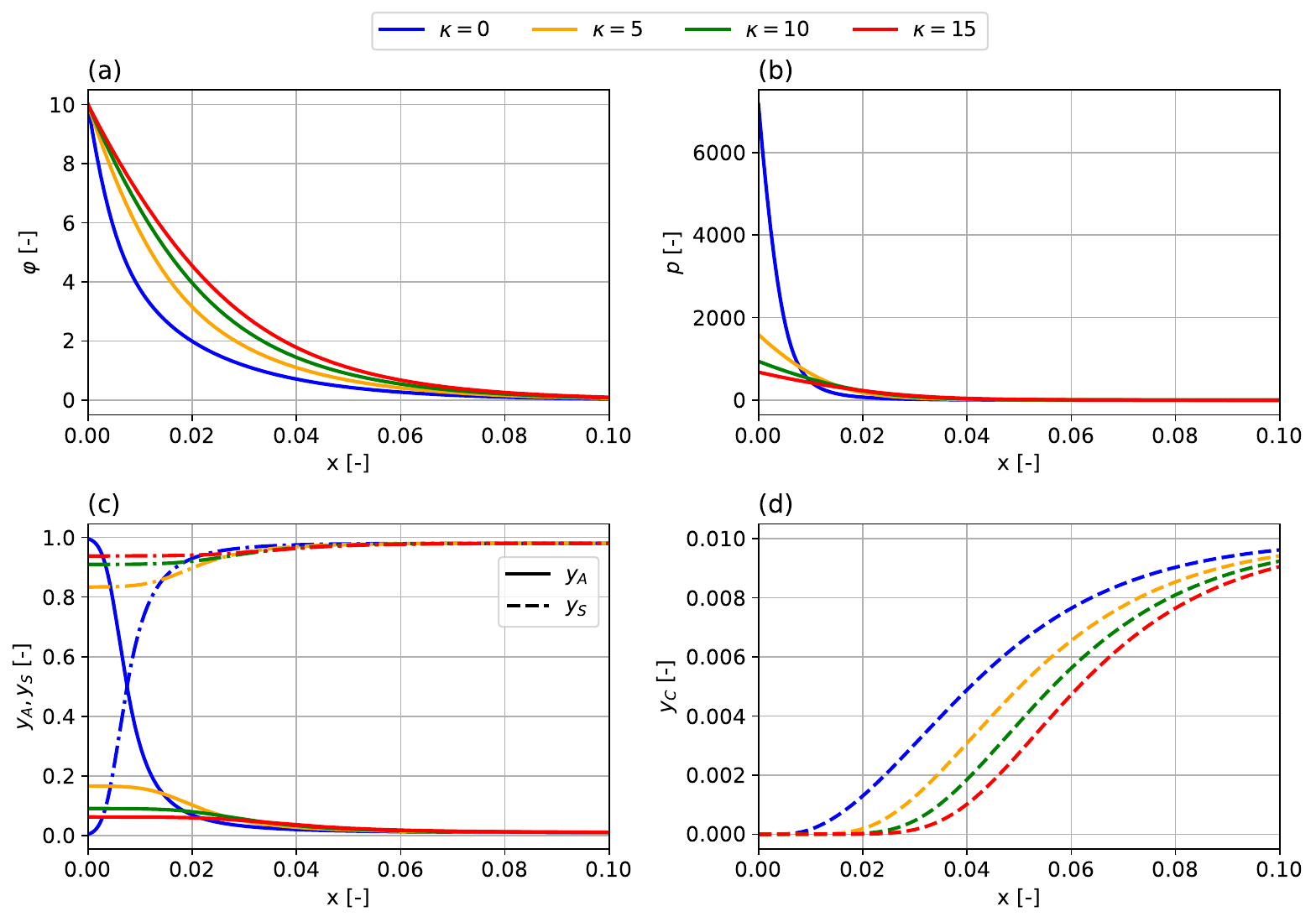}
	\caption{Effect of solvation number $\kappa$ on the dynamics of an incompressible ternary electrolyte: Profiles of (a) electric potential $\varphi$, (b) pressure $p$, (c) atomic fractions of anions ($y_A$) and neutral solvent ($y_S$), and (d) atomic fraction of cations ($y_C$) for four different solvation numbers $\kappa = 0$, 5, 10, and 15.}\label{Fig:8}
\end{figure}
%================================

\paragraph{Effect of solvation number}
%%%%%%%%%%%%%%%%%%%%%%%%%%%%%%%%%%%%%%%%%%%%%%%%%%%%%%%%%%

Figure~\ref{Fig:8} presents the impact of the solvation number $\kappa$ on the electric potential, pressure, and atomic fractions in an incompressible ternary electrolyte. Simulations are carried out for four values of the solvation number: $\kappa = 0$, 5, 10, and 15. These values represent increasing solvation effects, where higher $\kappa$ indicates stronger clustering of solvent molecules around the ions.
In Fig.~\ref{Fig:8}(a), the electric potential $\varphi$ displays a steeper gradient near the electrode as the solvation number increases. This sharpening indicates that the potential drop occurs over a narrower region, intensifying the electric field close to the boundary. Simultaneously, the decay becomes more gradual further from the interface, resulting in a higher overall voltage retained within the electrolyte domain. These changes reflect the solvation-induced redistribution of ions, which alters the charge separation structure near the interface. The pressure distribution shown in Fig.~\ref{Fig:8}(b) is also sensitive to solvation. As $\kappa$ increases, the pressure profile becomes broader and flatter, indicating an extended solvation region. The domain over which the pressure remains non-zero expands with increasing $\kappa$, highlighting the volumetric effects of solvated species, particularly near the electrode.

Fig.~\ref{Fig:8}(c)-(d) show the evolution of atomic fractions. The anion fraction $y_A$ exhibits sharp peaks near the interface for low $\kappa$ but saturates more gradually as solvation increases. With no solvation ($\kappa = 0$), the model allows $y_A$ to approach unity at the interface, but this is physically unrealistic. As $\kappa$ increases, steric hindrance from solvated ions limits this growth, leading to lower saturation levels and a wider solvation region. The cation profile $y_C$ displays similar behavior, with earlier saturation and more gradual transitions for larger solvation numbers. Meanwhile, the neutral solvent $y_S$ shows increasing concentration near the electrode for larger $\kappa$, compensating for the reduced ionic content due to solvation-induced blocking.
Physically, these results are consistent with the expectation that solvation prevents the electrolyte from being fully dominated by a single ionic species near the interface. In the absence of solvation, a positively charged electrode would attract a dense layer of anions, possibly leading to nonphysical oversaturation. With increasing solvation number, solvent molecules cluster around the ions, occupying volume and limiting excessive accumulation. As a result, cation and anion concentrations are suppressed near the interface, while the neutral solvent fraction increases.

%================================
%================================
\begin{figure}
	\centering
	\includegraphics[width=1.0\linewidth]{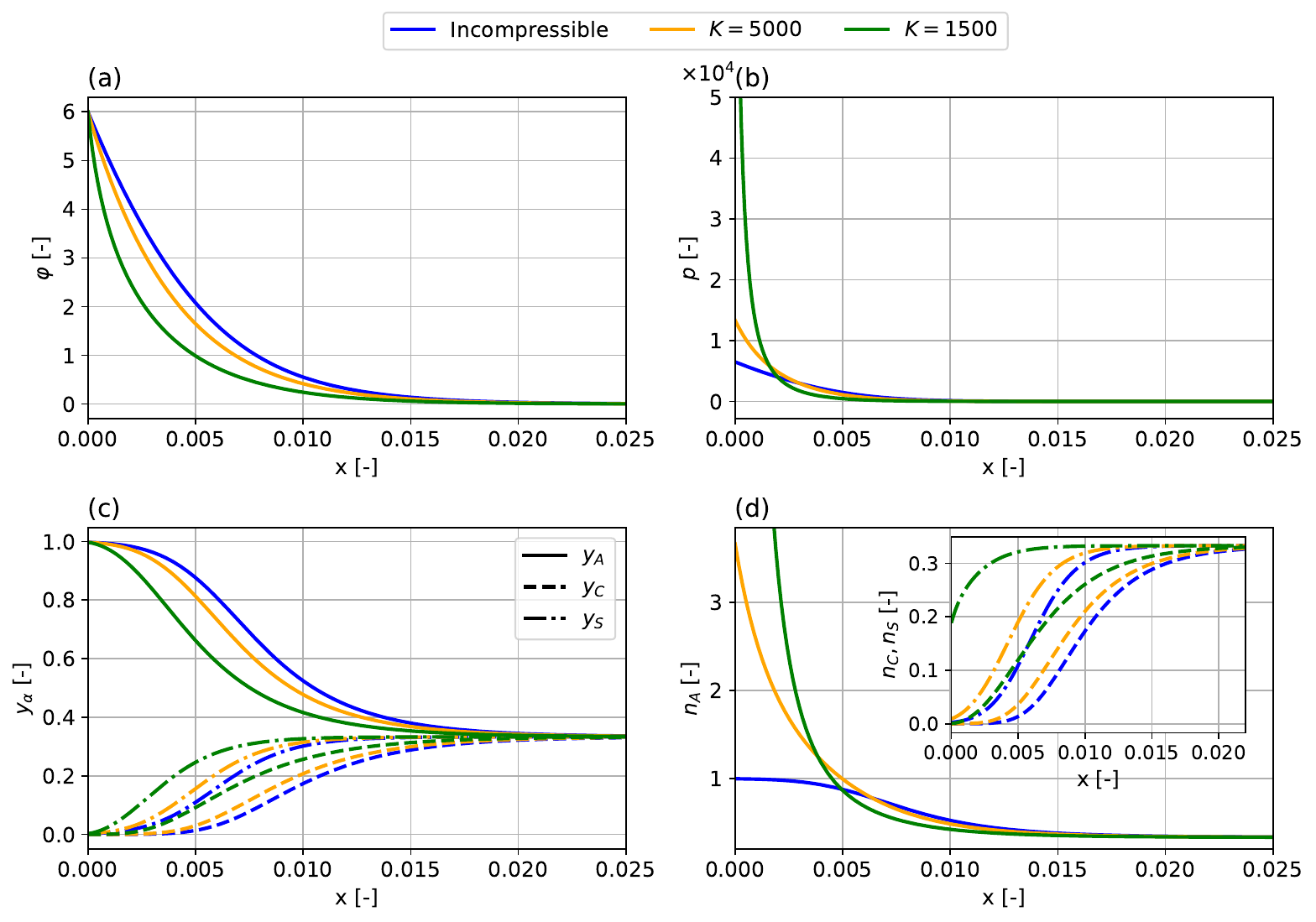}
	\caption{Effect of the bulk modulus on the dynamics of a ternary electrolyte: Profiles of (a) electric potential $\varphi$, (b) pressure $p$, (c) concentration (anions $y_{A}$, cations $y_{C}$, and neutral solvent $y_{S}$), (d) number density $n_{A}$ with $K\rightarrow \infty$ (incompressible), $K=5000$ and $K=1500$.}\label{Fig:9}
\end{figure}
%================================

\paragraph{Effect of compressibility}
%%%%%%%%%%%%%%%%%%%%%%%%%%%%%%%%%%%%%%%%%%%%%%%%%%%%%%%%%%

Figure~\ref{Fig:9} illustrates the impact of the dimensionless bulk modulus $K$ on the electrochemical behavior of a ternary electrolyte. Three cases are considered: an incompressible mixture ($K \rightarrow \infty$), and compressible mixtures with $K = 5000$ and $K = 1500$.
As shown in Fig.~\ref{Fig:9}(a), the electric potential distribution becomes more gradual and ``bulbous'' with decreasing bulk modulus, indicating increased compressibility. In the incompressible case, the steep potential gradient near the interface reflects strong electrostatic screening, while in compressible systems, the softening of the material allows the potential to decay more smoothly across a broader region. Fig.~\ref{Fig:9}(b) demonstrates that pressure decreases sharply near the interface for all values of $K$, but the pressure peak becomes more pronounced and spatially confined as the mixture becomes less compressible. In contrast, the incompressible case exhibits a broader distribution with lower peak magnitude, consistent with the constraint of constant volume.

In Fig.~\ref{Fig:9}(c), the atomic fractions of all species remain bounded within the physically admissible range $[0,1]$. As compressibility increases, the spatial region where saturation occurs (i.e., where $y_A \approx 1$ and $y_C \approx 0$) becomes narrower, reflecting a thinner electrical double layer. This narrowing indicates that higher compressibility reduces the effective crowding of ions near the electrode, leading to steeper gradients.
Fig.~\ref{Fig:9}(d) highlights a crucial difference between compressible and incompressible models: the total number density is no longer constant when $K$ is finite. While the incompressible model assumes uniform total density across the domain (hence atomic fractions directly scale to number densities), compressible systems allow local accumulation of particles. This leads to a sharp, exponential increase in $n_A$ near the electrode for lower values of $K$, driven by the applied potential. The inset confirms similar trends for $n_C$, though its magnitude remains lower due to electrostatic repulsion.

%%%%%%%%%%%%%%%%%%%%%%%%%%%%%%%%%%%%%%%%%%%%%%%%%%%%%%%%%%
\subsection{Differential double layer capacitance for a ternary electrolyte}\label{Sec:5.2}

The differential double layer capacitance is a key electrochemical property that characterizes the charge storage behavior of electrolyte systems near charged interfaces. In this section, we consider a ternary electrolyte system composed of monovalent anions and cations, with identical but opposite charges, i.e., $z_A = -1$, $z_C = +1$, and a neutral solvent with $z_S = 0$. The total electrostatic charge stored in the system is defined as the integral of the space charge density over the spatial domain:
%----------------------------------
\begin{equation}
	\label{Eq:39}
	Q(\varphi^L) = \int_\Omega n^F \, \text{d}\Omega,
\end{equation}
%----------------------------------
where $\varphi^L$ is the applied electric potential at the left boundary and $n_F$ is the dimensionless free space charge density.

The differential double layer capacitance $C_{\text{dl}}$ is then defined as the derivative of the total stored charge with respect to the applied boundary potential:
%--------------------------------
\begin{equation}
	\label{Eq:40}
	C_{\text{dl}} = \frac{\text{d}Q}{\text{d}\varphi^L} = \frac{\text{d}}{\text{d}\varphi^L} \int_\Omega n^F \, \text{d}\Omega.
\end{equation}
%--------------------------------

To systematically evaluate this, we impose Dirichlet boundary conditions for the atomic fractions of the ionic species (anions and cations) on the right side of the domain based on a specified molarity $M$:
\begin{equation}
	y_\alpha^R = \frac{M}{n^{\text{ref}}}, \quad \alpha \in \{A, C\},
\end{equation}
where $n^{\text{ref}}$ is the reference number density of the system.
For the simulations presented in this section, we use the default values for the scaled Debye length $\lambda^2$ and the pressure scaling factor $a^2$, as introduced earlier in the manuscript. The double layer capacitance is calculated with finite differences with respect to the applied boundary potential.

%================================
%================================
\begin{figure}
	\centering
	\includegraphics[width=1.0\linewidth]{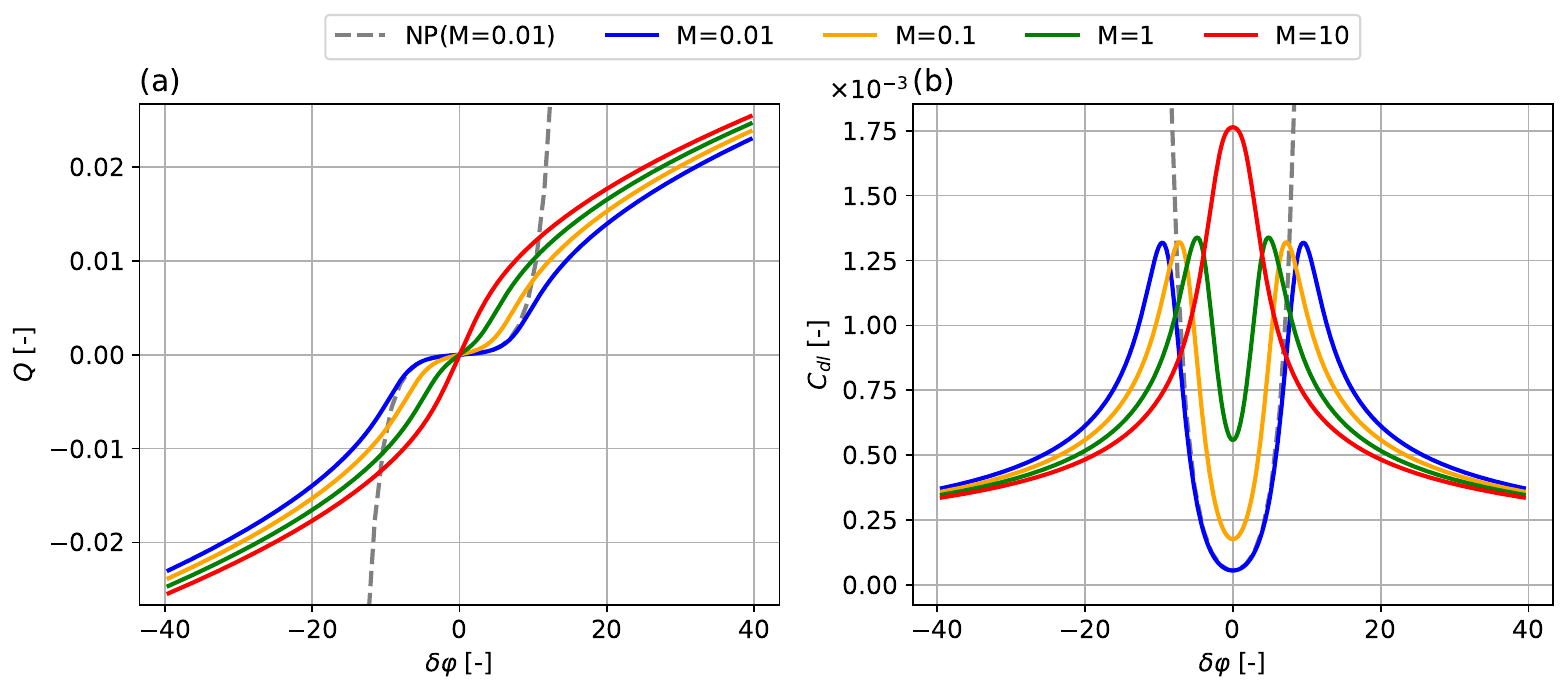}
	\caption{Impact of electrolyte molarity on one-dimensional differential double layer capacitance and comparison to classical NP model: profiles of (a) total electrostatic charge $Q$, and (b) differential double layer capacitance $C_{\text{dl}}$ with different molarity $M=0.01, 0.1, 1, 10$  $mol / L$.}\label{Fig:10}
\end{figure}
%================================

%%%%%%%%%%%%%%%%%%%%%%%%%%%%%%%%%%%%%%%%%%%%%%%%%%%%%%%%%%
\subsubsection{Effect of electrolyte molarity}\label{Sec:5.2.1}

Figure~\ref{Fig:10} illustrates the influence of electrolyte molarity on the total electrostatic charge $Q$ and the differential double layer capacitance $C_{\text{dl}}$ in a symmetric ternary electrolyte system. The molarity values considered are $M = 0.01$, $0.1$, $1.0$, $10.0$ mol/L, with results compared against the classical NP model. The applied potential difference $\delta \varphi$ is varied over a wide range to capture the full charge response of the system. As shown in Fig.~\ref{Fig:10}(a), the total charge $Q$ remains zero at $\delta \varphi = 0$ for all molarities, reflecting a globally charge-neutral system in the absence of an external field. As the potential difference increases, $Q$ rises monotonically in absolute value due to ion migration toward the oppositely charged electrode. For low molarities, a noticeable inflection or ``saddle point'' appears around $\delta \varphi = 0$, which is smoothed out and eventually vanishes for higher concentrations. At high molarity (e.g., $M = 10$ mol/L), the ionic population is large enough that the charge distribution evolves more smoothly with the applied field, leading to a more linear charge-voltage relationship in the central region. Fig.~\ref{Fig:10}(b) shows the corresponding differential capacitance curves. The capacitance is strictly positive and symmetric with respect to $\delta \varphi = 0$ for all cases. For low molarities, the capacitance exhibits a characteristic \textit{camel-shaped} profile: two local maxima appear at non-zero values of $\delta \varphi$, separated by a local minimum at the origin. These peaks occur where the ion concentrations approach their physical saturation limits (blocking values), which enhances space charge and thus capacitance.

As molarity increases, these two local maxima move closer together and eventually merge. For high molarity ($M = 10$ mol/L), the capacitance profile transitions to a \textit{bell-shaped} curve, with a single maximum at $\delta \varphi = 0$ and monotonic decay on either side. This change in shape reflects the increased availability of ions in the system and the delayed onset of saturation effects, requiring higher voltages to reach blocking concentrations. In all cases, the capacitance decays toward an asymptotic value as $|\delta \varphi|$ increases, due to saturation of ionic concentrations and reduction in the system's ability to further store charge.
The classical NP model, shown for comparison, fails to capture these physically meaningful behaviors. Both $Q$ and $C_{\text{dl}}$ diverge as the applied potential grows in magnitude, with $Q$ trending toward infinity or minus infinity depending on the sign of $\delta \varphi$. This nonphysical result underscores the limitations of the NP model in concentrated regimes and demonstrates the need for thermodynamically consistent formulations to properly capture charge saturation and steric effects in real systems.

%================================
%================================
\begin{figure}
	\centering
	\includegraphics[width=1.0\linewidth]{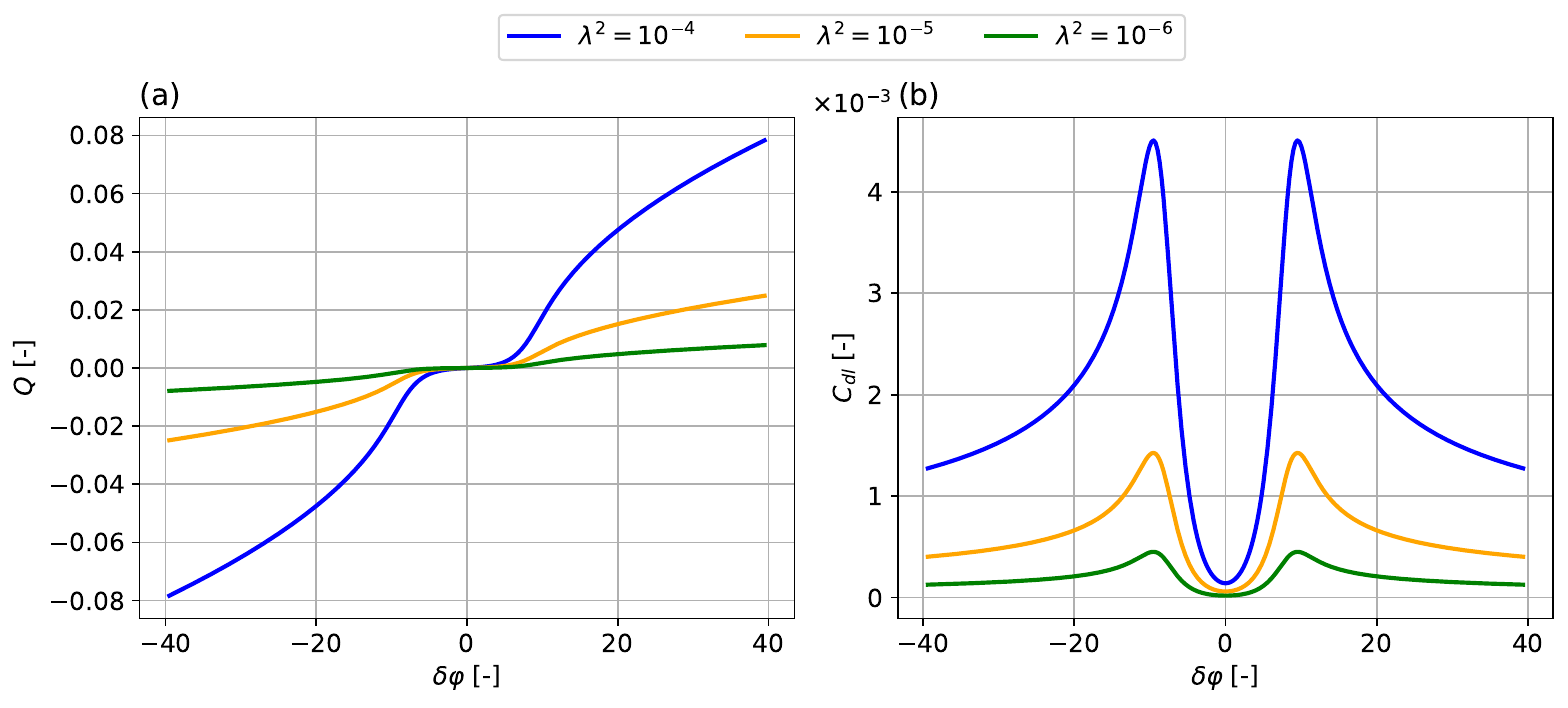}
	\caption{Impact of scaled Debye length on one-dimensional differential double layer capacitance: profiles of (a) total electrostatic charge $Q$, and (b) differential double layer capacitance $C_{\text{dl}}$ with different  Debye length $\lambda^{2}=10^{-4}, 10^{-5}, 10^{-6}$.}\label{Fig:11}
\end{figure}
%================================

%%%%%%%%%%%%%%%%%%%%%%%%%%%%%%%%%%%%%%%%%%%%%%%%%%%%%%%%%%
\subsubsection{Effect of scaled Debye length}\label{Sec:5.2.2}

Figure~\ref{Fig:11} explores the influence of the scaled Debye length $\lambda^2$ on total electrostatic charge $Q$ and differential double layer capacitance $C_{\text{dl}}$ in a one-dimensional symmetric ternary electrolyte. Three values of scaled Debye lengths are considered: $\lambda^{2}=10^{-4}, 10^{-5}, 10^{-6}$ with a fixed molarity of $M = 0.01\,\text{mol}/\text{L}$. As shown in Fig.~\ref{Fig:11}(a), the total charge $Q$ increases monotonically with the applied potential difference $\delta \varphi$ for all values of $\lambda^2$. Larger Debye lengths correspond to more extensive electric double layers, allowing greater charge separation and, thus, a higher net electrostatic charge stored in the system. This manifests as a steeper slope and larger maximum values of $Q$ for increasing $\lambda^2$. Physically, increasing $\lambda^2$ corresponds to a decrease in the reference number density or an increase in temperature or dielectric susceptibility, all of which enhance charge screening over a wider region. Fig.~\ref{Fig:11}(b) reveals the corresponding behavior of the differential capacitance $C_{\text{dl}}$. For all cases, the capacitance exhibits a characteristic camel-shaped curve, with two local maxima at finite values of $\delta \varphi$ and a local minimum at $\delta \varphi = 0$. As $\lambda^2$ increases, both the height and width of the capacitance profile increase, reflecting the system's enhanced ability to store charge across a broader spatial region. The scaling behavior of $C_{\text{dl}}$ with $\lambda^2$ is directly linked to the space charge distribution $n_F$, which in turn depends on the gradients in atomic fractions of the ionic species. A larger scaled Debye length results in more gradual ionic gradients, broader double layers, and more diffuse but extensive space charge regions. These properties collectively lead to higher differential capacitance values.

%================================
%================================
\begin{figure}
	\centering
	\includegraphics[width=1.0\linewidth]{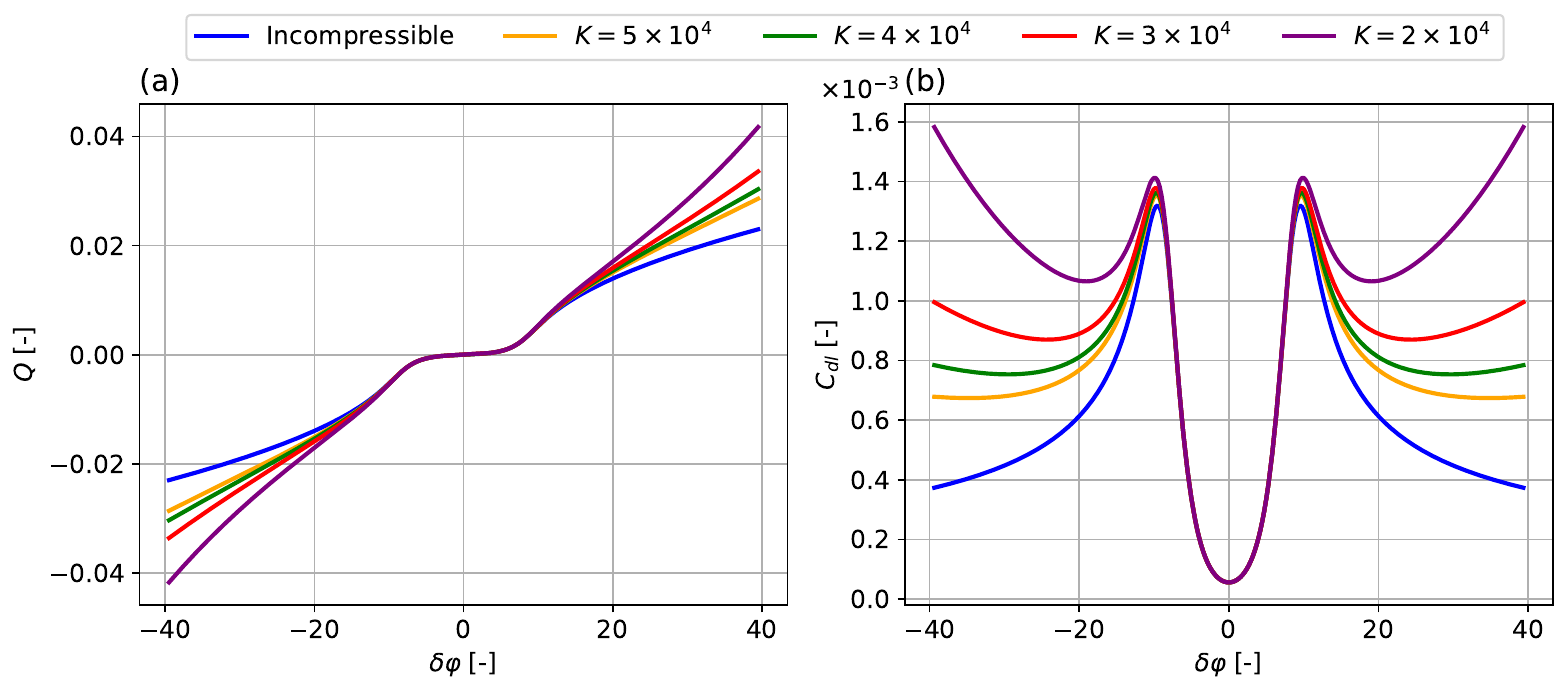}
	\caption{Impact of bulk modulus on one-dimensional differential double layer capacitance: profiles of (a) total electrostatic charge $Q$, and (b) differential double layer capacitance $C_{\text{dl}}$ with $K\rightarrow \infty$ (incompressible), $K=5 \times 10^{4}, 4 \times 10^{4}, 3 \times 10^{4}$, $2 \times 10^{4}$.}\label{Fig:12}
\end{figure}
%================================

%%%%%%%%%%%%%%%%%%%%%%%%%%%%%%%%%%%%%%%%%%%%%%%%%%%%%%%%%%
\subsubsection{Effect of bulk modulus}\label{Sec:5.2.3}

Figure~\ref{Fig:12} illustrates the effect of the bulk modulus $K$---which controls the compressibility of the electrolyte---on the one-dimensional electrostatic charge and differential double layer capacitance. The simulations are carried out for increasing compressibility, corresponding to decreasing $K$ values: $K=5 \times 10^{4}, 4 \times 10^{4}, 3 \times 10^{4}$, $2 \times 10^{4}$, with the incompressible limit ($K \to \infty$). All cases assume a fixed low molarity of $M = 0.01\,\text{mol}/\text{L}$. Fig.~\ref{Fig:12}(a) shows that the total electrostatic charge $Q$ increases monotonically with the applied potential difference $\delta \varphi$ for all values of $K$. As expected, compressibility affects the system's ability to accumulate space charge: lower values of $K$ (i.e., higher compressibility) lead to slightly higher stored charge at the same potential. The differences are small near $\delta \varphi = 0$, but they grow with increasing magnitude of the applied potential. As $K$ increases, the results converge smoothly toward the incompressible solution, confirming the theoretical expectation. The more pronounced effect of compressibility appears in the capacitance behavior, shown in Fig.~\ref{Fig:12}(b). For the incompressible case and high $K$ values, the capacitance exhibits a distinct \textit{camel-shaped} profile with two symmetric peaks and a local minimum at $\delta \varphi = 0$. This shape reflects the steric saturation of ions near the electrodes at high field strengths. However, as compressibility increases (i.e., with lower $K$), the profile evolves: the twin humps become broader and less pronounced, and for sufficiently low $K$ (e.g., $K = 2 \times 10^{4}$), the capacitance begins to increase again beyond the classical humps, deviating from the camel shape. This suggests a shift in charge accumulation behavior due to enhanced volume change in the electrolyte, allowing additional ionic accommodation and thus boosting capacitance at high fields. Physically, these trends arise because compressibility directly influences the local number density and, hence, the space charge distribution. Unlike the incompressible case, where the number density is constant, compressible systems allow local volume expansion, enabling more ions to be packed near the interface under strong fields. This flexibility affects both the total charge stored and the rate at which it changes with applied voltage.

%================================
%================================
\begin{figure}
	\centering
	\includegraphics[width=1.0\linewidth]{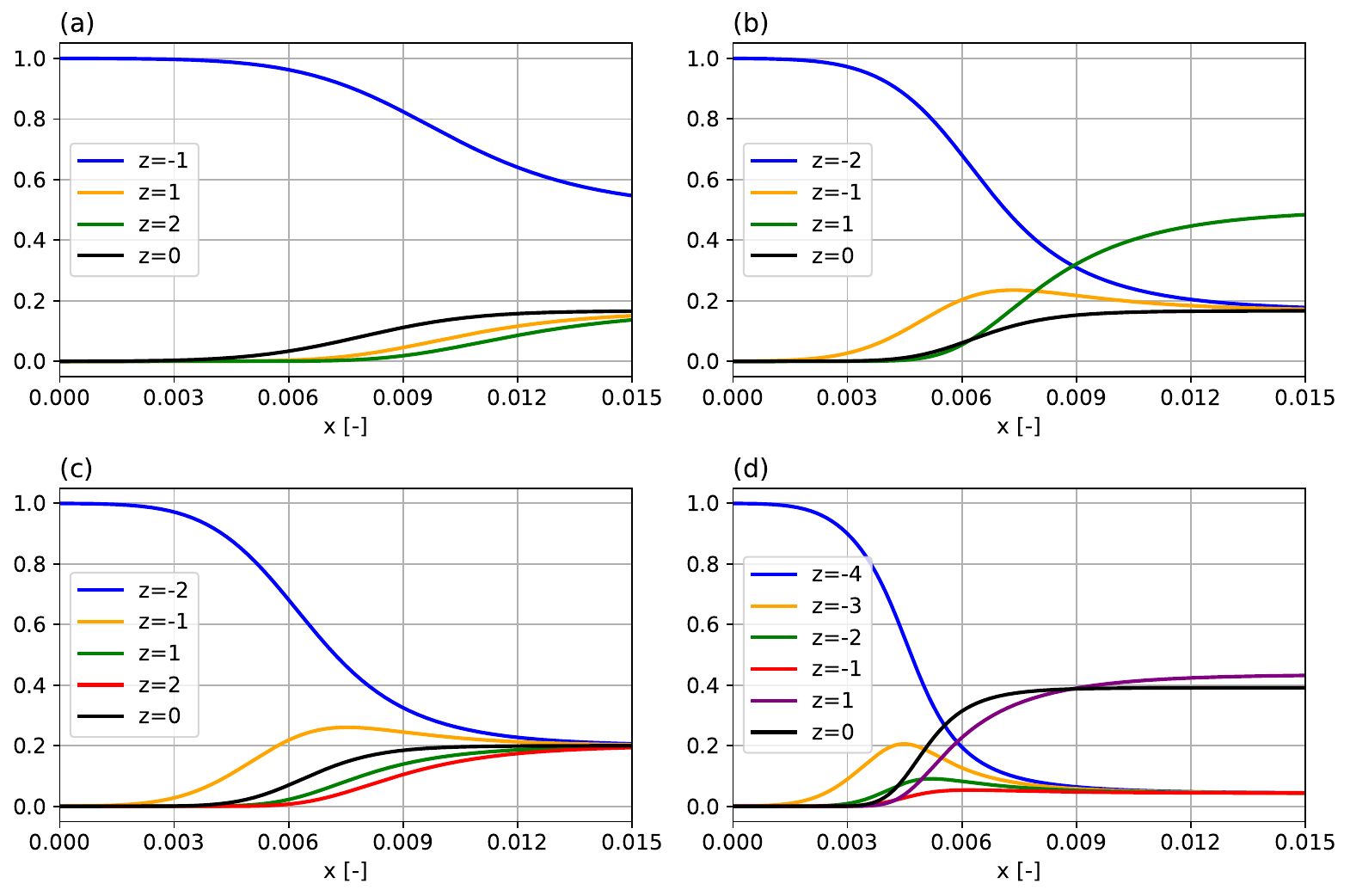}
	\caption{Comparison of four different mixtures with more than three constituents.}\label{Fig:13}
\end{figure}
%================================

%%%%%%%%%%%%%%%%%%%%%%%%%%%%%%%%%%%%%%%%%%%%%%%%%%%%%%%%%%
\subsection{Many constituent mixtures}\label{Sec:5.3}

Electrolyte mixtures comprising more than three species introduce additional physical and numerical complexity. In such systems, the total space charge density \( n_F \) is calculated as:
\begin{equation}
	n^F = n \sum_{\alpha=1}^{N} z_\alpha y_\alpha,
	\label{eq:space_charge_density}
\end{equation}
where \( N \neq 3 \). This expansion necessitates solving \( N - 1 \) partial mass balance equations in addition to the equations for electric potential and pressure, resulting in a total of \( N + 1 \) coupled partial differential equations.

To illustrate the behavior of systems with increasing ionic complexity, we consider four representative one-dimensional mixtures: two with four ionic species (Cases A and B), one with five species (Case C), and one with six species (Case D). Each mixture includes a neutral solvent species with charge \( z_N = 0 \), and all systems are assumed to be incompressible (i.e., \( K \to \infty \)) and in equilibrium, with vanishing ionic fluxes \( \boldsymbol{J}_\alpha = 0 \). A fixed potential difference \( \delta\varphi = 10 \) is applied across the domain, and the bulk pressure is set to zero (\( p^R = 0 \)).

The charge numbers for the constituent species in each case are defined as follows:
%----------------------------------------
\begin{align*}
	\text{Case A:} \quad & z^{(A)} = [-1, 1, 2], \\
	\text{Case B:} \quad & z^{(B)} = [-2, -1, 1], \\
	\text{Case C:} \quad & z^{(C)} = [-2, -1, 1, 2], \\
	\text{Case D:} \quad & z^{(D)} = [-4, -3, -2, -1, 1].
\end{align*}
%----------------------------------------

% In all cases, the solvent species is electrically neutral, i.e., \( z_N = 0 \).
The atomic fractions at the right boundary are prescribed for each species to ensure electroneutrality and realistic bulk composition. The boundary fractions are specified as:
%------------------------------
\begin{align}
	y_{1}^{R,(A)} & =\frac{3}{6},\quad y_{2}^{R,(A)}=\frac{1}{6},\quad y_{3}^{R,(A)}=\frac{1}{6},\\
	y_{1}^{R,(B)} & =\frac{1}{6},\quad y_{2}^{R,(B)}=\frac{1}{6},\quad y_{3}^{R,(B)}=\frac{3}{6},\\
	y_{1}^{R,(C)} & =y_{2}^{R,(C)}=y_{3}^{R,(C)}=y_{4}^{R,(C)}=y_{5}^{R,(C)}=\frac{1}{5},\\
	y_{1}^{R,(D)} & =y_{2}^{R,(D)}=y_{3}^{R,(D)}=y_{4}^{R,(D)}=\frac{1}{23},\quad y_{5}^{R,(D)}=\frac{10}{23},
\end{align}
%------------------------------
and $y_{N}^{R,(\cdot)}=1-\sum_{\alpha=1}^{N-1}y_{\alpha}^{R,(\cdot)}$.
The values for the fractions at the right boundary are chosen to have
charge neutrality in the bulk.

%%%%%%%%%%%%%%%%%%%%%%%%%%%%%%%%%%%%%%%%%%%%5

Figure~\ref{Fig:13} presents the atomic fraction distributions for four distinct electrolyte mixtures with more than three constituents. In each subplot---(a) through (d)---the behavior of ionic species is influenced by their respective charge numbers.
Across all mixtures, the atomic fractions of positively charged cations and the neutral solvent decrease monotonically toward zero near the boundary. Notably, the rate at which these fractions vanish increases with the charge number: species with higher positive charges diminish more rapidly.
For the negatively charged species, a more complex pattern emerges. Initially, all negatively charged species experience an increase in concentration; however, only the species with the largest negative charge number continues rising until it saturates near unity. Other anionic species exhibit a peak followed by a decline toward zero. The prominence and width of this local maximum diminish with decreasing charge magnitude, resulting in smaller saturation regions for species with smaller (less negative) charges. The pressure profile in each case mimics the qualitative behavior seen in simpler ternary systems but is scaled approximately by a factor proportional to the absolute value of the smallest (most negative) charge number in the mixture. The scaling is not exact across all mixtures, as reflected by slight deviations in the peak values. This discrepancy is likely due to the differing total charge number sums across mixtures, suggesting that the overall ionic charge composition also significantly influences the pressure response.

%================================
%================================
\begin{figure}
	\centering
	\includegraphics[width=0.7\linewidth]{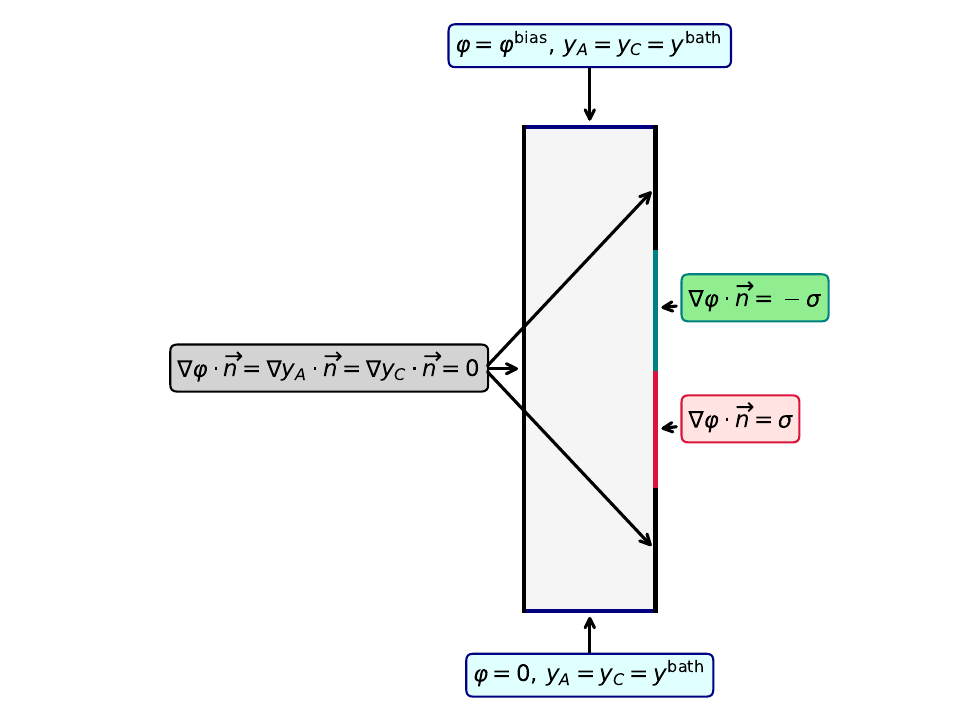}
	\caption{Boundary conditions for the electrolytic diode with $\varphi^{\text{bias}}\in [-10, 0, 10]$, $y_{A}^{\text{bath}}=y_{C}^{\text{bath}}=0.01$, $\sigma=350$. No Dirichlet boundary conditions and homogeneous Neumann boundary conditions are applied to the whole domain for the pressure.}\label{Fig:14}
\end{figure}
%================================

%%%%%%%%%%%%%%%%%%%%%%%%%%%%%%%%%%%%%%%%%%%%%%%%%%%%%%%%%%
\subsection{Two-dimensional electrolyte diode}\label{Sec:5.4}

Finally, we investigate a two-dimensional electrolytic diode consisting of two ion reservoirs connected by a nanofluidic channel~\cite{fuhrmann2016numerical}. The reservoirs contain predefined concentrations of anions and cations, and the channel includes a spatially varying surface charge distribution implemented via Neumann boundary conditions on the electric potential.
%---------------------------------------
The setup involves a ternary electrolyte within the rectangular domain $\Omega = [0, 0.02] \times [0, 0.1]$, with model parameters: scaled Debye length $\lambda^{2}=8.553\cdot 10^{-6}$, pressure scaling factor  $a^{2}=7.5412\cdot 10^{-4}$, and solvation number $\kappa=5$. At the top and bottom boundaries \((x, y) = (x, 0)\) and \((x, y) = (x, 0.1)\), the atomic fractions are fixed as \(y_A = y_C = y_\text{bath} = 0.01\). The electric potential is set to \(\varphi = 0\) at the lower boundary and \(\varphi = \varphi^\text{bias}\) at the upper boundary.

A nonuniform surface charge is applied along the right vertical boundary \((x = 0.02)\) using the Neumann condition:
\begin{equation}
	\nabla\varphi\cdot\overrightarrow{n}=g_{\varphi}=\begin{cases}
		-\sigma & \text{for } x=0.02,0.025\leq y<0.05,\\
		\sigma & \text{for } x=0.02,0.05\leq y\leq0.075,\\
		0 & \text{else},
	\end{cases}
\end{equation}
%--------------------------------
with \(\sigma = 350\). Figure~\ref{Fig:14} provides a visual schematic of these boundary conditions. This configuration establishes an internal electric field gradient that mimics diode-like behavior---preferentially permitting ion flow in one direction depending on the polarity of \(\varphi^\text{bias}\).

%================================
%================================
\begin{figure}
	\centering
	\includegraphics[width=1.0\linewidth]{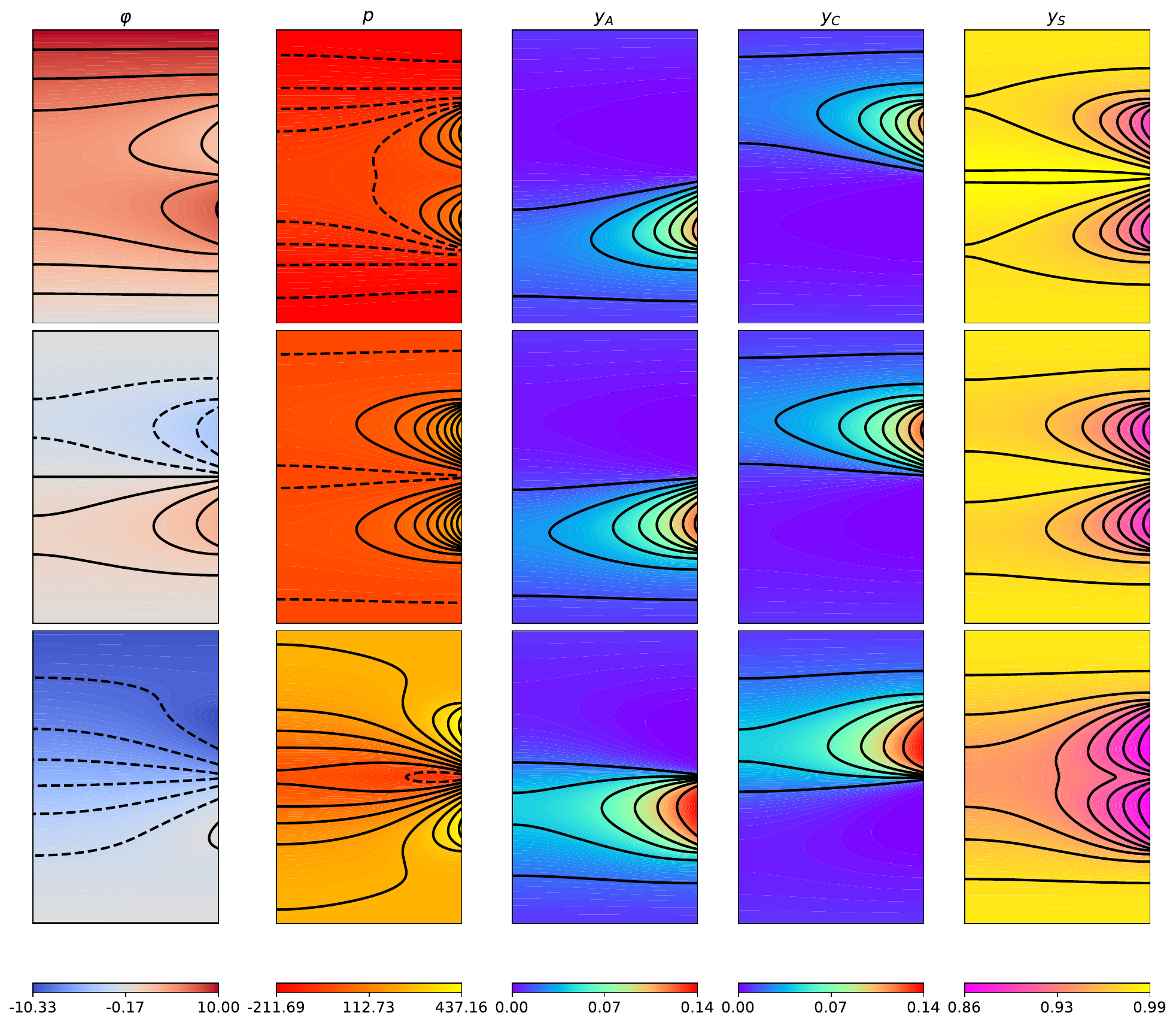}
	\caption{Comparison of a forward bias (top, $\varphi^{\text{bias}}=10$),
		no bias (center, $\varphi^{\text{bias}}=0$) and a backward bias (bottom,
		$\varphi^{\text{bias}}=-10$) with the scaled Debye length $\lambda^{2}=8.553 \cdot 10^{-6}$, the pressure scaling factor $a^{^{2}}=7.5412 \cdot 10^{-4}$, a solvation number $\kappa=5$ and a bath concentration for the anions and cations
		of 0.01.}\label{Fig:15}
\end{figure}
%================================

Figure~\ref{Fig:15} presents simulation results for a two-dimensional electrolyte diode under three different bias conditions: forward bias (\(\varphi^{\text{bias}} = 10\)), no bias (\(\varphi^{\text{bias}} = 0\)), and backward bias (\(\varphi^{\text{bias}} = -10\)). The columns show contour plots of the electric potential \(\varphi\), pressure \(p\), anion concentration \(y_A\), cation concentration \(y_C\), and neutral solvent concentration \(y_S\).
In the forward bias case, the electric potential increases from its minimum at the bottom boundary (\(y = 0\)) to a global maximum at the top boundary (\(y = 0.1\)). Due to the spatially varying Neumann boundary conditions on the right-hand wall, the potential exhibits local extrema---specifically, one local maximum and one local minimum---before reaching the global maximum. The pressure, in contrast, is compressed toward the center of the domain, forming a high-pressure zone between the regions of applied surface charge, while the top and bottom edges exhibit comparatively low pressures. The ionic species (anions and cations) are driven toward the outer regions of the domain, creating a spatial separation that reflects rectification behavior.
Under zero bias, the electric potential achieves symmetric extrema near the regions of surface charge application, and the overall potential distribution appears balanced. The pressure field, while still structured, now exhibits multiple local minima: at the top, bottom, and center of the domain. The ion concentrations remain relatively evenly distributed, with minimal drift toward either boundary, and only minor diffusion effects visible.

For the backward bias configuration, the electric potential increases from the bottom, reaches a peak, sharply drops to a minimum, and then increases again toward the top boundary. This non-monotonic behavior mirrors the reversed driving force and highlights the influence of internal surface charges. Correspondingly, the pressure profile exhibits two strong peaks, separated by a sharp minimum, and eventually decreases to a lower value at the top boundary. The ionic concentrations respond by slightly shifting inward, concentrating closer to the centerline of the domain, unlike the outward drift seen in the forward bias case.
Interestingly, despite these distinct electrostatic and pressure variations, the ionic concentrations do not show a strong rectification effect in terms of amplitude when comparing the three cases. This agrees with the findings of Fuhrmann~\cite{fuhrmann2016numerical}, where qualitative trends match well despite differences in domain geometry and parameter choices. However, as shown in Gaudeul et al.~\cite{gaudeul2022entropy}, such rectification effects can emerge if differences in specific volume or solvation numbers are pronounced enough. In this simulation, a solvation number \(\kappa = 5\) was used uniformly for ions, and the neutral solvent's specific volume was not altered, which explains the absence of a clear rectification signature in ion concentrations.

%%%%%%%%%%%%%%%%%%%%%%%%%%%%%%%%%%%%%%%%%%%%%%%%%%%%%%%%%%%
%%%%%%%%%%%%%%%%%%%%%%%%%%%%%%%%%%%%%%%%%%%%%%%%%%%%%%%%%%
\section{Concluding remarks and outlook}\label{Sec:6}

In this study, we presented a finite element solver for a thermodynamically consistent electrolyte model that rigorously captures the complex interplay between ionic transport, electrostatic interactions, and pressure dynamics in multicomponent electrolyte mixtures. The model is grounded in the principles of non-equilibrium thermodynamics and ensures strict adherence to mass conservation, charge neutrality, and entropy production. Unlike classical models such as the Nernst--Planck system, the framework captures steric saturation (via ideal mixing and volume-fraction bounds), solvation-induced volume changes, and pressure coupling; explicit hard-core diameters and nonlocal packing correlations are not included. Despite the model's increased complexity, the numerical framework exhibits excellent convergence and stability properties, supported by a modular variational formulation that is well-suited for further extensions.

The implementation of the model in the FEniCSx computing platform demonstrates both robustness and flexibility, enabling the simulation of one- and two-dimensional systems under a wide range of boundary conditions and configurations. Validation against benchmark problems and comparisons with classical models confirmed the improved physical fidelity of the solver, particularly in regimes involving high ionic concentrations or strong electrochemical gradients. The solver successfully handled systems with multiple ionic constituents, capturing the emergent phenomena arising from asymmetries in ion valence and spatially varying boundary conditions. Simulation results revealed key features of electrolyte behavior, including electric double layer formation, rectification phenomena, and the influence of parameters such as the solvation number, Debye length, and compressibility.

The aim of present study to develop a finite element solver that provides a versatile and physically consistent tool for simulating multicomponent electrolyte systems across diverse applications such as energy storage, water purification, biological transport, and microfluidics. Future work will focus on extending the model to include fluid flow for electrokinetic effects, temperature gradients for thermos-diffusion, and electrochemical reactions at interfaces. Further improvements include an explicit treatment of ion sizes and the incorporation of nonlocal correlations. To recover such effects, one may employ more sophisticated theories, e.g., classical DFT with hard-sphere mixtures and weighted-density approximations~\cite{mier-y-teranNonlocalFreeenergyDensityfunctional1990}, or integral-equation approaches for finite-size ions, such as HNC/MSA~\cite{lozada-cassouApplicationHypernettedChain1982}, which can yield oscillatory layering and even charge inversion~\cite{grebergChargeInversionElectric1998}; see also modified Poisson--Boltzmann formulations with finite-size corrections~\cite{outhwaiteImprovedModifiedPoisson1983}. Enhancing computational performance through adaptive meshing and integrating the solver into multi-physics frameworks will further broaden its applicability to complex, coupled electrochemical systems.

%%%%%%%%%%%%%%%%%%%%%%%%%%%%%%%%%%%%%%%%%%%%%%%%%%%%%%%%%%%
%%%%%%%%%%%%%%%%%%%%%%%%%%%%%%%%%%%%%%%%%%%%%%%%%%%%%%%%%%
\section*{Acknowledgments}
S.S., S.B., and M.T. acknowledge the support by the German Research Foundation (DFG), within the Research Unit FOR5409 under project numbers 463312734 and 517510462. L.T. and M.T. also acknowledge the support by the German Research Foundation (DFG), project B04 (504291427) within the SFB 1481 (442047500).

%%%%%%%%%%%%%%%%%%%%%%%%%%%%%%%%%%%%%%%%%%%%%%%%%%%%%%%%%%%
%%%%%%%%%%%%%%%%%%%%%%%%%%%%%%%%%%%%%%%%%%%%%%%%%%%%%%%%%%
\section*{Data availability \& reproducibility}
We provide a reproducibility repository~\cite{Habscheid_2025} containing the source code, scripts, and data to reproduce the results from Section~\ref{Sec:4} and Section~\ref{Sec:5}.

%%%%%%%%%%%%%%%%%%%%%%%%%%%%%%%%%%%%%%%%%%%%%%%%%%%%%%%%%%
% \section*{Data availability}
% We provide a reproducibility repository on Gitlab, which is archived on Zenodo, that contains data and code.

%%%%%%%%%%%%%%%%%%%%%%%%%%%%%%%%%%%%%%%%%%%%%%%%%%%%%%%%%%%
%%%%%%%%%%%%%%%%%%%%%%%%%%%%%%%%%%%%%%%%%%%%%%%%%%%%%%%%%%
\appendix
\section{Appendix: Nomenclature}

% \section*{\textcolor{red}{Appendix: }Nomenclature}

\begin{longtable}{ll}
	\toprule
	\textbf{Symbol} & \textbf{Description (with units or values)} \\
	\midrule
	$e_0$            & Elementary charge [$1.602 \times 10^{-19}$ C] \\
	$\varepsilon_0$  & Vacuum permittivity [$8.85 \times 10^{-12}$ F/m] \\
	$\mu_0$          & Magnetic permeability of vacuum [$4\pi \times 10^{-7}$ N/A$^2$] \\
	$k$              & Boltzmann constant [$1.381 \times 10^{-23}$ J/K] \\
	$N_A$            & Avogadro constant [$6.022 \times 10^{23}$ 1/mol] \\
	\\
	$n_\alpha$       & Number density of species $\alpha$ [1/m$^3$] \\
	$y_\alpha$       & Atomic (mole) fraction of species $\alpha$ [-] \\
	$m_\alpha$       & Mass of species $\alpha$ [kg] \\
	$z_\alpha$       & Charge number of species $\alpha$ [-] \\
	$\boldsymbol{J}_\alpha$ & Diffusion flux of species $\alpha$ [kg/m$^2$/s] \\
	$g_\alpha$ & Specific Gibbs energy of species $\alpha$ [J/kg] \\
	$\mu_\alpha$     & Chemical potential of species $\alpha$ [J/kg] \\
	$\kappa_\alpha$  & Solvation number of species $\alpha$ [-] \\
	$\boldsymbol{v}_\alpha$ & Velocity of species $\alpha$ [m/s] \\
	$\boldsymbol{u}_\alpha$ & Diffusion velocity of species $\alpha$ [m/s] \\
	$\rho_\alpha$    & Partial mass density of species $\alpha$ [kg/m$^3$] \\
	$V_\alpha$       & Specific volume of species $\alpha$ [m$^3$] \\
	\\
	$\varphi$           & Electrostatic potential [V] \\
	$p$              & Pressure [Pa] \\
	$n$              & Total number density [1/m$^3$] \\
	$T$              & Absolute temperature [K] \\
	% $n^{\text{ref}}$ & Reference number density [1/m$^3$] \\
	\\
	$K$              & Bulk modulus [Pa] \\
	$M$				& Molarity [mol/m$^3$] \\
	$\boldsymbol{E}$     & Electric field [V/m] \\
	% $\boldsymbol{B}$     & Magnetic field [T] \\
	$n^e$            & Total charge [As/m$^3$] \\
	$n^F$            & Free (space) charge [As/m$^3$] \\
	$n^P$            & Polarized charge [As/m$^3$] \\
	% $\boldsymbol{j}^e$   & Total current density [A/m$^2$] \\
	% $\boldsymbol{j}^F$   & Free current density [A/m$^2$] \\
	% $\boldsymbol{j}^P$   & Polarized current density [A/m$^2$] \\
	$\boldsymbol{P}$     & Polarization vector [C/m$^2$] \\
	$\chi$           & Dielectric susceptibility [-] \\
	$\boldsymbol{M}_{\alpha\beta}$ & Mobility (kinetic) matrix [kg·K·s/m$^3$] \\
	$\boldsymbol{v}$     & Barycentric velocity [m/s] \\
	$\rho$           & Total mass density [kg/m$^3$] \\
	$\rho \boldsymbol{v}$ & Total momentum [kg/m$^2$/s] \\
	$\boldsymbol{\sigma}$ & Cauchy stress tensor [Pa] \\
	% $\rho \boldsymbol{b}$ & Gravitational force density [kg/m$^2$/s$^2$] \\
	% $\boldsymbol{k}$     & Lorentz force [kg/m$^2$/s$^2$] \\
	\\
	$(\cdot)^{\text{ref}}$ & Reference state of quantity \\
	$(\cdot)^L$, $(\cdot)^R$ & Value at left/right boundary \\
	$\delta(\cdot) = (\cdot)^L - (\cdot)^R$ & Jump between boundaries \\
	$(\cdot)_\alpha$ & Property of species $\alpha$ \\
	\bottomrule
\end{longtable}\label{app1}

%%%%%%%%%%%%%%%%%%%%%%%%%%%%%%%%%%%%%%%%%%%%%%%%%%%%%%%%%%%

%%%%%%%%%%%%%%%%%%%%%%%%%%%%%%%%%%%%%%%%%%%%%%%%%%%%%%%%%%%%%%%%%%%%%%%%%%%%%%%
%%%%%%%%%%%%%%%%%%%%%%%%%%%%%%%%%%%%%%%%%%%%%%%%%%%%%%%%%%%%%%%%%%%%%%%%%%%%%%%
\printcredits{}

% %% Loading bibliography style file
% \bibliographystyle{model1-num-names}
% \bibliographystyle{cas-model2-names}

% \bibliography{cas-refs}

% ignore underful warnings in the bibliography: hard to fix
% LTeX: enabled=true
\apptocmd{\sloppy}{\hbadness10000\relax}{}{}
% LTeX: enabled=false

%=====================================
% \bibliographystyle{plain}  % or another style like 'unsrt', 'alpha', etc.
% \bibliographystyle{elsearticle-num}
\bibliographystyle{plainurl}
\bibliography{cas-refs}   % refers to references.bib

\begin{thebibliography}{10}

\bibitem{anderko2002electrolyte}
A.~Anderko, P.~Wang, and M.~Rafal.
\newblock Electrolyte solutions: From thermodynamic and transport property
  models to the simulation of industrial processes.
\newblock {\em Fluid Phase Equilibria}, 194:123--142, 2002.
\newblock \href {https://doi.org/10.1016/S0378-3812(01)00645-8}
  {\path{doi:10.1016/S0378-3812(01)00645-8}}.

\bibitem{anderson2010capacitive}
M.~A. Anderson, A.~L. Cudero, and J.~Palma.
\newblock Capacitive deionization as an electrochemical means of saving energy
  and delivering clean water. comparison to present desalination practices:
  Will it compete?
\newblock {\em Electrochimica Acta}, 55(12):3845--3856, 2010.
\newblock \href {https://doi.org/10.1016/j.electacta.2010.02.012}
  {\path{doi:10.1016/j.electacta.2010.02.012}}.

\bibitem{Ankur2025electrolyte}
Ankur, R.~Jiwari, and S.~Singh.
\newblock Finite element method for the numerical simulation of modified
  poisson--nernst--planck/navier--stokes model.
\newblock {\em arXiv preprint arXiv:2409.08746}, 2025.
\newblock Revision submitted to Applied Numerical Mathematics.
\newblock \href {https://doi.org/10.48550/arXiv.2409.08746}
  {\path{doi:10.48550/arXiv.2409.08746}}.

\bibitem{aseyev2014electrolytes}
G.~G. Aseyev.
\newblock {\em Electrolytes: Supramolecular Interactions and Non-Equilibrium
  Phenomena in Concentrated Solutions}.
\newblock CRC Press, 2014.
\newblock \href {https://doi.org/10.1201/b17818} {\path{doi:10.1201/b17818}}.

\bibitem{BarattaEtal2023}
Igor~A. Baratta, Joseph~P. Dean, J{\o}rgen~S. Dokken, Michal Habera, Jack~S.
  Hale, Chris~N. Richardson, Marie~E. Rognes, Matthew~W. Scroggs, Nathan Sime,
  and Garth~N. Wells.
\newblock {DOLFINx}: the next generation {FEniCS} problem solving environment.
\newblock preprint, 2023.
\newblock \href {https://doi.org/10.5281/zenodo.10447666}
  {\path{doi:10.5281/zenodo.10447666}}.

\bibitem{braun2015thermodynamically}
S.~Braun, C.~Yada, and A.~Latz.
\newblock Thermodynamically consistent model for space-charge-layer formation
  in a solid electrolyte.
\newblock {\em Journal of Physical Chemistry C}, 119:22281--22288, 2015.
\newblock \href {https://doi.org/10.1021/acs.jpcc.5b02679}
  {\path{doi:10.1021/acs.jpcc.5b02679}}.

\bibitem{de2013non}
S.~R. De~Groot and P.~Mazur.
\newblock {\em Non-Equilibrium Thermodynamics}.
\newblock Courier Corporation, 2013.
\newblock URL: \url{https://books.google.de/books?id=mfFyG9jfaMYC}.

\bibitem{dingSecondorderPositiveUnconditional2024}
Jie Ding and Shenggao Zhou.
\newblock Second-order, positive, and unconditional energy dissipative scheme
  for modified {{Poisson}}--{{Nernst}}--{{Planck}} equations.
\newblock {\em Journal of Computational Physics}, 510:113094, 2024.
\newblock \href {https://doi.org/10.1016/j.jcp.2024.113094}
  {\path{doi:10.1016/j.jcp.2024.113094}}.

\bibitem{dreyer2014mixture}
W.~Dreyer, C.~Guhlke, and M.~Landstorfer.
\newblock A mixture theory of electrolytes containing solvation effects.
\newblock {\em Electrochemistry Communications}, 43:75--78, 2014.
\newblock \href {https://doi.org/10.1016/j.elecom.2014.03.015}
  {\path{doi:10.1016/j.elecom.2014.03.015}}.

\bibitem{dreyer2013overcoming}
W.~Dreyer, C.~Guhlke, and R.~M{\"u}ller.
\newblock Overcoming the shortcomings of the {Nernst--Planck} model.
\newblock {\em Physical Chemistry Chemical Physics}, 15:7075--7086, 2013.
\newblock \href {https://doi.org/10.1039/C3CP44390F}
  {\path{doi:10.1039/C3CP44390F}}.

\bibitem{dreyer2018bulk}
W.~Dreyer, C.~Guhlke, and R.~M{\"u}ller.
\newblock Bulk-surface electrothermodynamics and applications to
  electrochemistry.
\newblock {\em Entropy}, 20:939, 2018.
\newblock \href {https://doi.org/10.3390/e20120939}
  {\path{doi:10.3390/e20120939}}.

\bibitem{Fuhrmann_LiquidElectrolytes_GitHub}
J.~Fuhrman.
\newblock {LiquidElectrolytes.jl}.
\newblock \url{https://github.com/j-fu/LiquidElectrolytes.jl}, 2024.

\bibitem{Fuhrmann2015}
J.~Fuhrmann.
\newblock Comparison and numerical treatment of generalised {Nernst--Planck}
  models.
\newblock {\em Computer Physics Communications}, 196:166--178, 2015.
\newblock \href {https://doi.org/10.1016/j.cpc.2015.06.004}
  {\path{doi:10.1016/j.cpc.2015.06.004}}.

\bibitem{fuhrmann2016numerical}
J.~Fuhrmann.
\newblock A numerical strategy for {Nernst--Planck} systems with solvation
  effect.
\newblock {\em Fuel Cells}, 16:704--714, 2016.
\newblock \href {https://doi.org/10.1002/fuce.201500215}
  {\path{doi:10.1002/fuce.201500215}}.

\bibitem{Gao2017}
H.~Gao and D.~He.
\newblock Linearized conservative finite element methods for the
  {Nernst--Planck--Poisson} equations.
\newblock {\em Journal of Scientific Computing}, 72:1269--1289, 2017.
\newblock \href {https://doi.org/10.1007/s10915-017-0394-y}
  {\path{doi:10.1007/s10915-017-0394-y}}.

\bibitem{Gao2018}
H.~Gao and P.~Sun.
\newblock A linearized local conservative mixed finite element method for
  {Poisson--Nernst--Planck} equations.
\newblock {\em Journal of Scientific Computing}, 77:793--817, 2018.
\newblock \href {https://doi.org/10.1007/s10915-018-0727-5}
  {\path{doi:10.1007/s10915-018-0727-5}}.

\bibitem{gaudeul2022entropy}
B.~Gaudeul and J.~Fuhrmann.
\newblock Entropy and convergence analysis for two finite volume schemes for a
  {Nernst--Planck--Poisson} system with ion volume constraints.
\newblock {\em Numerische Mathematik}, 151:99--149, 2022.
\newblock \href {https://doi.org/10.1007/s00211-022-01279-y}
  {\path{doi:10.1007/s00211-022-01279-y}}.

\bibitem{grebergChargeInversionElectric1998}
Hans Greberg and Roland Kjellander.
\newblock Charge inversion in electric double layers and effects of different
  sizes for counterions and coions.
\newblock {\em J. Chem. Phys.}, 108(7):2940--2953, 1998.
\newblock \href {https://doi.org/10.1063/1.475681}
  {\path{doi:10.1063/1.475681}}.

\bibitem{grmela1997dynamics}
M.~Grmela and H.~C. {\"O}ttinger.
\newblock Dynamics and thermodynamics of complex fluids. {I}. {D}evelopment of
  a general formalism.
\newblock {\em Physical Review E}, 56:6620, 1997.
\newblock \href {https://doi.org/10.1103/PhysRevE.56.6620}
  {\path{doi:10.1103/PhysRevE.56.6620}}.

\bibitem{Habscheid_2025}
J.~Habscheid, S.~Singh, L.~Theisen, S.~Braun, and M.~Torrilhon.
\newblock fxdgm: A nonlinear, mixed finite element solver for the dgm
  electrolyte model, 2025.
\newblock \href {https://doi.org/10.5281/zenodo.16948003}
  {\path{doi:10.5281/zenodo.16948003}}.

\bibitem{He2017}
M.~He and P.~Sun.
\newblock Error analysis of mixed finite element method for
  {Poisson-Nernst-Planck} system.
\newblock {\em Numerical Methods for Partial Differential Equations},
  33:1924--1948, 2017.
\newblock \href {https://doi.org/10.1002/num.22170}
  {\path{doi:10.1002/num.22170}}.

\bibitem{He2018}
M.~He and P.~Sun.
\newblock Mixed finite element analysis for the
  {Poisson--Nernst--Planck/Stokes} coupling.
\newblock {\em Journal of Computational and Applied Mathematics}, 341:61--79,
  2018.
\newblock \href {https://doi.org/10.1016/j.cam.2018.04.003}
  {\path{doi:10.1016/j.cam.2018.04.003}}.

\bibitem{Mingyan}
M.~He and P.~Sun.
\newblock Mixed finite element method for modified
  {Poisson--Nernst--Planck/Navier--Stokes} equations.
\newblock {\em Journal of Scientific Computing}, 8(3):1--33, 2021.
\newblock \href {https://doi.org/10.1007/s10915-021-01478-z}
  {\path{doi:10.1007/s10915-021-01478-z}}.

\bibitem{horng2012pnp}
T.~L. Horng, T.~C. Lin, C.~Liu, and B.~Eisenberg.
\newblock {PNP} equations with steric effects: A model of ion flow through
  channels.
\newblock {\em Journal of Physical Chemistry B}, 116:11422--11441, 2013.
\newblock \href {https://doi.org/10.1021/jp305273n}
  {\path{doi:10.1021/jp305273n}}.

\bibitem{janek2016solid}
J.~Janek and W.~G. Zeier.
\newblock A solid future for battery development.
\newblock {\em Nature Energy}, 1(9):1--4, 2016.
\newblock \href {https://doi.org/10.1038/nenergy.2016.141}
  {\path{doi:10.1038/nenergy.2016.141}}.

\bibitem{Jerome1985}
J.~W. Jerome.
\newblock Consistency of semiconductor modeling: An existence/stability
  analysis for the stationary {Van Boosbroeck} system.
\newblock {\em SIAM Journal on Applied Mathematics}, 45:565--590, 1985.
\newblock \href {https://doi.org/10.1137/0145034} {\path{doi:10.1137/0145034}}.

\bibitem{Jerome2002}
J.~W. Jerome.
\newblock Analytical approaches to charge transport in a moving medium.
\newblock {\em Transport Theory and Statistical Physics}, 31:333--366, 2002.
\newblock \href {https://doi.org/10.1081/TT-120015505}
  {\path{doi:10.1081/TT-120015505}}.

\bibitem{jerome2012analysis}
J.~W. Jerome.
\newblock {\em Analysis of Charge Transport: A Mathematical Study of
  Semiconductor Devices}.
\newblock Springer Berlin, Heidelberg, 2012.
\newblock \href {https://doi.org/10.1007/978-3-642-79987-7}
  {\path{doi:10.1007/978-3-642-79987-7}}.

\bibitem{Jerome1991}
J.~W. Jerome and T.~Kerkhoven.
\newblock A finite element approximation theory for the drift diffusion
  semiconductor model.
\newblock {\em SIAM Journal on Numerical Analysis}, 28(2):403--422, 1991.
\newblock \href {https://doi.org/10.1137/0728023} {\path{doi:10.1137/0728023}}.

\bibitem{kilic2007steric}
M.~S. Kilic, M.~Z. Bazant, and A.~Ajdari.
\newblock Steric effects in the dynamics of electrolytes at large applied
  voltages. {I}. {D}ouble-layer charging.
\newblock {\em Physical Review E}, 75:021502, 2007.
\newblock \href {https://doi.org/10.1103/PhysRevE.75.021502}
  {\path{doi:10.1103/PhysRevE.75.021502}}.

\bibitem{Kim2022}
S.~Kim, M.~A. Khanwalea, R.~K. Anand, and B.~Ganapathysubramanian.
\newblock Computational framework for resolving boundary layers in
  electrochemical systems using weak imposition of {Dirichlet} boundary
  conditions.
\newblock {\em Finite Elements in Analysis and Design}, 205:103749, 2022.
\newblock \href {https://doi.org/10.1016/j.finel.2022.103749}
  {\path{doi:10.1016/j.finel.2022.103749}}.

\bibitem{Kornyshev1981Conductivity}
A.~A. Kornyshev and M.~A. Vorotyntsev.
\newblock Conductivity and space charge phenomena in solid electrolytes with
  one mobile charge carrier species, a review with original material.
\newblock {\em Electrochimica Acta}, 26(3):303--323, 1981.
\newblock \href {https://doi.org/10.1016/0013-4686(81)85017-7}
  {\path{doi:10.1016/0013-4686(81)85017-7}}.

\bibitem{landstorferTheoryStructureMetalelectrolyte2016}
M.~Landstorfer, C.~Guhlke, and W.~Dreyer.
\newblock Theory and structure of the metal-electrolyte interface incorporating
  adsorption and solvation effects.
\newblock {\em Electrochimica Acta}, 201:187--219, 2016.
\newblock \href {https://doi.org/10.1016/j.electacta.2016.03.013}
  {\path{doi:10.1016/j.electacta.2016.03.013}}.

\bibitem{Linga2020}
G.~Linga, A.~Bolet, and J.~Mathiesen.
\newblock Transient electrohydrodynamic flow with concentration-dependent fluid
  properties: Modelling and energy-stable numerical schemes.
\newblock {\em Journal of Computational Physics}, 412:109430, 2020.
\newblock \href {https://doi.org/10.1016/j.jcp.2020.109430}
  {\path{doi:10.1016/j.jcp.2020.109430}}.

\bibitem{liu2017free}
H.~Liu and Z.~Wang.
\newblock A free energy satisfying discontinuous galerkin method for
  one-dimensional poisson--nernst--planck systems.
\newblock {\em Journal of Computational Physics}, 328:413--437, 2017.
\newblock \href {https://doi.org/10.1016/j.jcp.2016.10.008}
  {\path{doi:10.1016/j.jcp.2016.10.008}}.

\bibitem{liu2022positivity}
H.~Liu, Z.~Wang, P.~Yin, and H.~Yu.
\newblock Positivity-preserving third order dg schemes for
  poisson--nernst--planck equations.
\newblock {\em Journal of Computational Physics}, 452:110777, 2022.
\newblock \href {https://doi.org/10.1016/j.jcp.2021.1107776}
  {\path{doi:10.1016/j.jcp.2021.1107776}}.

\bibitem{logg2012finite}
A.~Logg, K.~A. Mardal, and G.~N. Wells.
\newblock {\em Automated Solution of Differential Equations by the Finite
  Element Method}.
\newblock Springer, 2012.
\newblock \href {https://doi.org/10.1007/978-3-642-23099-8}
  {\path{doi:10.1007/978-3-642-23099-8}}.

\bibitem{lotsch2017relevance}
B.~V. Lotsch and J.~Maier.
\newblock Relevance of solid electrolytes for lithium-based batteries: A
  realistic view.
\newblock {\em Journal of Electroceramics}, 38:128--141, 2017.
\newblock \href {https://doi.org/10.1007/s10832-017-0091-0}
  {\path{doi:10.1007/s10832-017-0091-0}}.

\bibitem{lozada-cassouApplicationHypernettedChain1982}
Marcelo Lozada-Cassou, Rafael Saavedra-Barrera, and Douglas Henderson.
\newblock The application of the hypernetted chain approximation to the
  electrical double layer: {{Comparison}} with {{Monte Carlo}} results for
  symmetric salts.
\newblock {\em J. Chem. Phys.}, 77(10):5150--5156, 1982.
\newblock \href {https://doi.org/10.1063/1.443691}
  {\path{doi:10.1063/1.443691}}.

\bibitem{luPoissonNernstPlanckEquationsSimulating2011}
Benzhuo Lu and Y.~C. Zhou.
\newblock Poisson-{{Nernst-Planck Equations}} for {{Simulating Biomolecular
  Diffusion-Reaction Processes II}}: {{Size Effects}} on {{Ionic
  Distributions}} and~{{Diffusion-Reaction Rates}}.
\newblock {\em Biophysical Journal}, 100(10):2475--2485, 2011.
\newblock \href {https://doi.org/10.1016/j.bpj.2011.03.059}
  {\path{doi:10.1016/j.bpj.2011.03.059}}.

\bibitem{Mansoori1971equilibrium}
G.~A. Mansoori, N.~F. Carnahan, K.~E. Starling, and T.~W. Leland, Jr.
\newblock Equilibrium thermodynamic properties of the mixture of hard spheres.
\newblock {\em Journal of Chemical Physics}, 54:1523, 1971.
\newblock \href {https://doi.org/10.1063/1.1675048}
  {\path{doi:10.1063/1.1675048}}.

\bibitem{mier-y-teranNonlocalFreeenergyDensityfunctional1990}
L.~Mier-y Teran, S.~H. Suh, H.~S. White, and H.~T. Davis.
\newblock A nonlocal free-energy density-functional approximation for the
  electrical double layer.
\newblock {\em J. Chem. Phys.}, 92(8):5087--5098, 1990.
\newblock \href {https://doi.org/10.1063/1.458542}
  {\path{doi:10.1063/1.458542}}.

\bibitem{Nernst1888}
W.~Nernst.
\newblock Zur kinetik der in l\"osung befindlichen k\"orper.
\newblock {\em Zeitschrift f\"ur Physikalische Chemie}, 2:613--637, 1888.
\newblock \href {https://doi.org/10.1515/zpch-1888-0274}
  {\path{doi:10.1515/zpch-1888-0274}}.

\bibitem{Nernst1889}
W.~Nernst.
\newblock Die elektromotorische wirksamkeit der ionen.
\newblock {\em Zeitschrift f\"ur Physikalische Chemie}, 4:129--181, 1889.
\newblock \href {https://doi.org/10.1515/zpch-1889-0412}
  {\path{doi:10.1515/zpch-1889-0412}}.

\bibitem{ottinger2005beyond}
H.~C. \"Ottinger.
\newblock {\em Beyond Equilibrium Thermodynamics}.
\newblock Wiley-Interscience, Hoboken, 2005.
\newblock \href {https://doi.org/10.1002/0471727903}
  {\path{doi:10.1002/0471727903}}.

\bibitem{outhwaiteImprovedModifiedPoisson1983}
Christopher~W. Outhwaite and Lutful~B. Bhuiyan.
\newblock An improved modified {{Poisson}}--{{Boltzmann}} equation in
  electric-double-layer theory.
\newblock {\em J. Chem. Soc., Faraday Trans. 2}, 79(5):707--718, 1983.
\newblock \href {https://doi.org/10.1039/F29837900707}
  {\path{doi:10.1039/F29837900707}}.

\bibitem{Park1997}
J.-H. Park and J.~W. Jerome.
\newblock Qualitative properties of steady-state {Poisson--Nernst--Planck}
  systems: Mathematical study.
\newblock {\em SIAM Journal on Applied Mathematics}, 57(3):609--630, 1997.
\newblock \href {https://doi.org/10.1137/S0036139995279809}
  {\path{doi:10.1137/S0036139995279809}}.

\bibitem{placke2017lithium}
T.~Placke, R.~Kloepsch, S.~D{\"u}hnen, and M.~Winter.
\newblock Lithium ion, lithium metal, and alternative rechargeable battery
  technologies: The odyssey for high energy density.
\newblock {\em Journal of Solid State Electrochemistry}, 21:1939--1964, 2017.
\newblock \href {https://doi.org/10.1007/s10008-017-3610-7}
  {\path{doi:10.1007/s10008-017-3610-7}}.

\bibitem{Planck1890one}
M.~Planck.
\newblock Ueber die erregung von electricit\"at und w\"arme in electrolyten.
\newblock {\em Annalen der Physik}, 275:561--576, 1890.
\newblock \href {https://doi.org/10.1002/andp.18902750202}
  {\path{doi:10.1002/andp.18902750202}}.

\bibitem{Planck1890two}
M.~Planck.
\newblock Ueber die potentialdifferenz zwischen zwei verd\"unnten l\"osungen
  bin\"arer electrolyte.
\newblock {\em Annalen der Physik}, 276:161--186, 1890.
\newblock \href {https://doi.org/10.1002/andp.18902760802}
  {\path{doi:10.1002/andp.18902760802}}.

\bibitem{Prohl2009}
A.~Prohl and M.~Schmuck.
\newblock Convergent discretizations for the {Nernst--Planck--Poisson} system.
\newblock {\em Numerische Mathematik}, 111:591--630, 2009.
\newblock \href {https://doi.org/10.1007/s00211-008-0194-2}
  {\path{doi:10.1007/s00211-008-0194-2}}.

\bibitem{Prohl2010}
A.~Prohl and M.~Schmuck.
\newblock Convergent finite element discretizations of the
  {Navier--Stokes--Nernst--Planck-Poisson} system.
\newblock {\em ESAIM: M2AN}, 44(3):531--571, 2010.
\newblock \href {https://doi.org/10.1051/m2an/2010013}
  {\path{doi:10.1051/m2an/2010013}}.

\bibitem{prohl2010convergent}
A.~Prohl and M.~Schmuck.
\newblock Convergent finite element discretizations of the
  {Navier-Stokes-Nernst-Planck-Poisson} system.
\newblock {\em ESAIM: M2AN}, 44:531--571, 2010.
\newblock \href {https://doi.org/10.1051/m2an/2010013}
  {\path{doi:10.1051/m2an/2010013}}.

\bibitem{Ray2012}
N.~Ray, A.~Muntean, and P.~Knabner.
\newblock Rigorous homogenization of a {Stokes--Nernst--Planck--Poisson}
  system.
\newblock {\em Journal of Mathematical Analysis and Applications},
  390:374--393, 2012.
\newblock \href {https://doi.org/10.1016/j.jmaa.2012.01.052}
  {\path{doi:10.1016/j.jmaa.2012.01.052}}.

\bibitem{roubivcek2006incompressible}
T.~Roub{\'\i}{\v{c}}ek.
\newblock Incompressible ionized fluid mixtures.
\newblock {\em Continuum Mechanics and Thermodynamics}, 17:493--509, 2006.
\newblock \href {https://doi.org/10.1007/s00161-006-0010-0}
  {\path{doi:10.1007/s00161-006-0010-0}}.

\bibitem{Schmuck2009}
M.~Schmuck.
\newblock Analysis of the {Navier--Stokes--Nernst--Planck--Poisson} system.
\newblock {\em Mathematical Models and Methods in Applied Sciences},
  19(6):993--1015, 2009.
\newblock \href {https://doi.org/10.1142/S0218202509003693}
  {\path{doi:10.1142/S0218202509003693}}.

\bibitem{Schmuck2011}
M.~Schmuck.
\newblock Modeling and deriving porous media {Stokes-Poisson-Nernst-Planck}
  equations by a multi-scale approach.
\newblock {\em Communications in Mathematical Sciences}, 9(3):685--710, 2011.
\newblock \href {https://doi.org/10.4310/CMS.2011.V9.N3.A3}
  {\path{doi:10.4310/CMS.2011.V9.N3.A3}}.

\bibitem{sparnaay1958corrections}
M.~J. Sparnaay.
\newblock Corrections of the theory of the flat diffuse double layer.
\newblock {\em Recueil des Travaux Chimiques des Pays-Bas}, 77(9):872--888,
  1958.
\newblock \href {https://doi.org/10.1002/recl.19580770911}
  {\path{doi:10.1002/recl.19580770911}}.

\bibitem{Sun2016}
Y.~Sun, P.~Sun, B.~Zheng, and G.~Lin.
\newblock Error analysis of finite element method for {Poisson--Nernst--Planck}
  equations.
\newblock {\em Journal of Computational and Applied Mathematics}, 301:28--43,
  2016.
\newblock \href {https://doi.org/10.1016/j.cam.2016.01.028}
  {\path{doi:10.1016/j.cam.2016.01.028}}.

\bibitem{suss2015water}
M.~E. Suss, S.~Porada, X.~Sun, P.~M. Biesheuvel, J.~Yoon, and V.~Presser.
\newblock Water desalination via capacitive deionization: What is it and what
  can we expect from it?
\newblock {\em Energy \& Environmental Science}, 8(8):2296--2319, 2015.
\newblock \href {https://doi.org/10.1039/C5EE00519A}
  {\path{doi:10.1039/C5EE00519A}}.

\bibitem{Tresset2008generalized}
G.~Tresset.
\newblock Generalized {Poisson--Fermi} formalism for investigating size
  correlation effects with multiple ions.
\newblock {\em Physical Review E}, 78(6):061506, 2008.
\newblock \href {https://doi.org/10.1103/PhysRevE.78.061506}
  {\path{doi:10.1103/PhysRevE.78.061506}}.

\bibitem{wright2007introduction}
M.~R. Wright.
\newblock {\em An Introduction to Aqueous Electrolyte Solutions}.
\newblock John Wiley \& Sons, 2007.
\newblock \href {https://doi.org/10.1002/cphc.200800219}
  {\path{doi:10.1002/cphc.200800219}}.

\bibitem{xia2017electrolytes}
L.~Xia, L.~Yu, D.~Hu, and G.~Z. Chen.
\newblock Electrolytes for electrochemical energy storage.
\newblock {\em Materials Chemistry Frontiers}, 1:584--618, 2017.
\newblock \href {https://doi.org/10.1039/C6QM00169F}
  {\path{doi:10.1039/C6QM00169F}}.

\bibitem{Xie2020}
D.~Xie and B.~Lu.
\newblock An effective finite element iterative solver for a
  {Poisson--Nernst--Planck} ion channel model with periodic boundary
  conditions.
\newblock {\em SIAM Journal on Scientific Computing}, 42(6):B1490--B1516, 2020.
\newblock \href {https://doi.org/10.1137/19M1297099}
  {\path{doi:10.1137/19M1297099}}.

\bibitem{Yang2013}
Y.~Yang and B.~Lu.
\newblock An error analysis for the finite element approximation to the
  steady-state {Poisson-Nernst--Planck} equations.
\newblock {\em Advances in Applied Mathematics and Mechanics}, 5(1):113--130,
  2013.
\newblock \href {https://doi.org/10.4208/aamm.11-m11184}
  {\path{doi:10.4208/aamm.11-m11184}}.

\bibitem{yang2011electrochemical}
Z.~Yang, J.~Zhang, M.~C.~W. Kintner-Meyer, X.~Lu, D.~Choi, J.~P. Lemmon, and
  J.~Liu.
\newblock Electrochemical energy storage for green grid.
\newblock {\em Chemical Reviews}, 111(5):3577--3613, 2011.
\newblock \href {https://doi.org/10.1021/cr100290v}
  {\path{doi:10.1021/cr100290v}}.

\bibitem{yue2016all}
L.~Yue, J.~Ma, J.~Zhang, J.~Zhao, S.~Dong, Z.~Liu, G.~Cui, and L.~Chen.
\newblock All solid-state polymer electrolytes for high-performance lithium ion
  batteries.
\newblock {\em Energy Storage Materials}, 5:139--164, 2016.
\newblock \href {https://doi.org/10.1016/j.ensm.2016.07.003}
  {\path{doi:10.1016/j.ensm.2016.07.003}}.

\end{thebibliography}

\end{document}